%% file: text7.tex
\documentclass[useAMS,usenatbib]{mn2e}
\usepackage{graphicx}
\def\apj{ApJ}
\def\apjs{ApJS}
\def\apjl{ApJL}
\def\aap{A\&A}
\def\aaps{A\&AS}

\def\mnras{MNRAS}

\def\pasp{PASP}
\def\nat{Nature}

\def\rmxaa{Revista Mexicana de Astronomia y Astrofisica}
\def\sci{Science}

\title[CSFGs with extremely high O$_{32}$]
{LBT observations of compact star-forming galaxies with 
extremely high [O {\sc iii}]/[O {\sc ii}] flux ratios: He {\sc i} emission-line ratios as diagnostics of Lyman continuum leakage}
\author[Y. I. Izotov et al.]{Y. I.\ Izotov$^{1}$,
T. X.\ Thuan$^{2}$ and N. G.\ Guseva$^{1}$ \\
                $^{1}$Main Astronomical Observatory,
                     Ukrainian National Academy of Sciences,
                     Zabolotnoho 27, Kyiv 03143,  Ukraine, 
                     \\ ~~izotov@mao.kiev.ua, guseva@mao.kiev.ua\\
                $^{2}$Astronomy Department, University of Virginia, 
                     P.O. Box 400325, Charlottesville, VA 22904-4325,
                     txt@virginia.edu\\
}
\begin{document}


\pagerange{\pageref{firstpage}--\pageref{lastpage}} \pubyear{2012}

\maketitle

\label{firstpage}

\begin{abstract}
We present Large Binocular Telescope spectrophotometric observations of 
five low-redshift ($z$ $<$ 0.070) compact star-forming galaxies (CSFGs) 
with extremely high emission-line ratios
O$_{32}$ = [O~{\sc iii}]$\lambda$5007/[O~{\sc ii}]$\lambda$3727, ranging from
23 to 43. Galaxies with such high O$_{32}$ are thought to be promising 
candidates for leaking large amounts of Lyman continuum (LyC) radiation and, at 
high redshifts, for contributing to the reionization of the Universe. 
The equivalent widths EW(H$\beta$) of the H$\beta$ emission line in the 
studied galaxies are very high, $\sim$ 350 -- 520\AA,
indicating very young ages for the star formation bursts, $<$ 3 Myr. All 
galaxies are characterized by low oxygen abundances 12+logO/H = 7.46 -- 7.79 
and low masses $M_\star$ $\sim$ 10$^6$ -- 10$^7$ M$_\odot$, much lower
than the  $M_\star$  for known low-redshift LyC leaking galaxies, but probably 
more typical of the hypothetical population of low-luminosity
dwarf LyC leakers at high redshifts. A broad H$\alpha$ emission line is 
detected in the spectra of all CSFGs, possibly related to expansion motions of 
supernova remnants. Such rapid ionized gas motions would facilitate the 
escape of the resonant Ly$\alpha$ emission from the galaxy. We show that 
high O$_{32}$ may not be a sufficient condition for LyC leakage and 
propose new diagnostics based on the He {\sc i} $\lambda$3889/$\lambda$6678 
and $\lambda$7065/$\lambda$6678 emission-line flux ratios. 
Using these diagnostics we find that three CSFGs
in our sample are likely to have density-bounded H~{\sc ii} regions and are 
thus leaking large amounts of LyC radiation. The amount of leaking LyC radiation
is probably much lower in the other two CSFGs. 
\end{abstract}


\section{Introduction}\label{sec:INT}

Recent studies of low-redshift compact star-forming galaxies (CSFGs) found
in the Sloan Digital Sky Survey (SDSS) demonstrated that they are low-mass and 
low-metallicity galaxies with star formation occuring in short bursts of 
a few million years duration
\citep*[e.g. ][]{Ca09,I11,I16c}. In the literature, and depending on their 
redshifts, subsets of these galaxies
have been variously called blue compact dwarf (BCD) 
galaxies at redshifts $z < 0.1$ \citep{LT86,T08} because of their blue colours,
``Green peas'' (GPs) which at redshifts of $\sim$ 0.1 -- 0.3 appear green on 
composite SDSS images \citep{Ca09} and luminous 
compact galaxies (LCGs) with diverse colours in the wider redshift range from 
0 to $\sim$0.6 \citep{I11}.
 
These properties together with the fact that CSFGs and high-redshift 
star-forming galaxies (SFGs) follow similar mass-metallicity and 
luminosity-metallicity relations suggest that CSFGs are good local counterparts 
of the Lyman-alpha emitting (LAE) galaxies and Lyman-break
galaxies (LBGs) \citep{I15}. The presence of 
strong emission lines in the optical spectra of the H~{\sc ii} regions of CSFGs,
powered by $\sim$ 10$^{3}$ -- 10$^{5}$ O-stars which produce plenty of 
ionizing radiation, make them promising candidates for 
leaking ionizing radiation into the intergalactic medium (IGM) 
\citep{Ca09,I11,JO13,S15}. 
It is indeed 
thought that dwarf galaxies with similar properties are responsible
for the reionization of the Universe at redshifts $z$ = 5 -- 10 
\citep{O09,WC09,M13,Y11,B15}. 

Many CSFGs are characterised by high line ratios O$_{32}$ = 
[O~{\sc iii}]$\lambda$5007/[O~{\sc ii}]$\lambda$3727, reaching
values of up to 60 in some galaxies \citep{S15}. Such high values
may indicate that their H~{\sc ii} regions are density-bounded, allowing escape
of ionizing radiation into the IGM, as suggested e.g.\ by \citet{JO13}, 
\citet{NO14} and \citet{N16}. However, that is not the 
only explanation. A high O$_{32}$
can also be caused by a low metallicity, a high ionization  parameter
or a hard  ionizing  radiation \citep{JO13,S15}. From 
various emission-line diagnostics, \citet{S15} reached the 
conclusion that CSFGs are in general opaque to 
ionizing radiation, and that they have 
low absolute escape fractions, not exceeding 10 percent.

\begin{figure*}
\includegraphics[angle=0,width=0.19\linewidth]{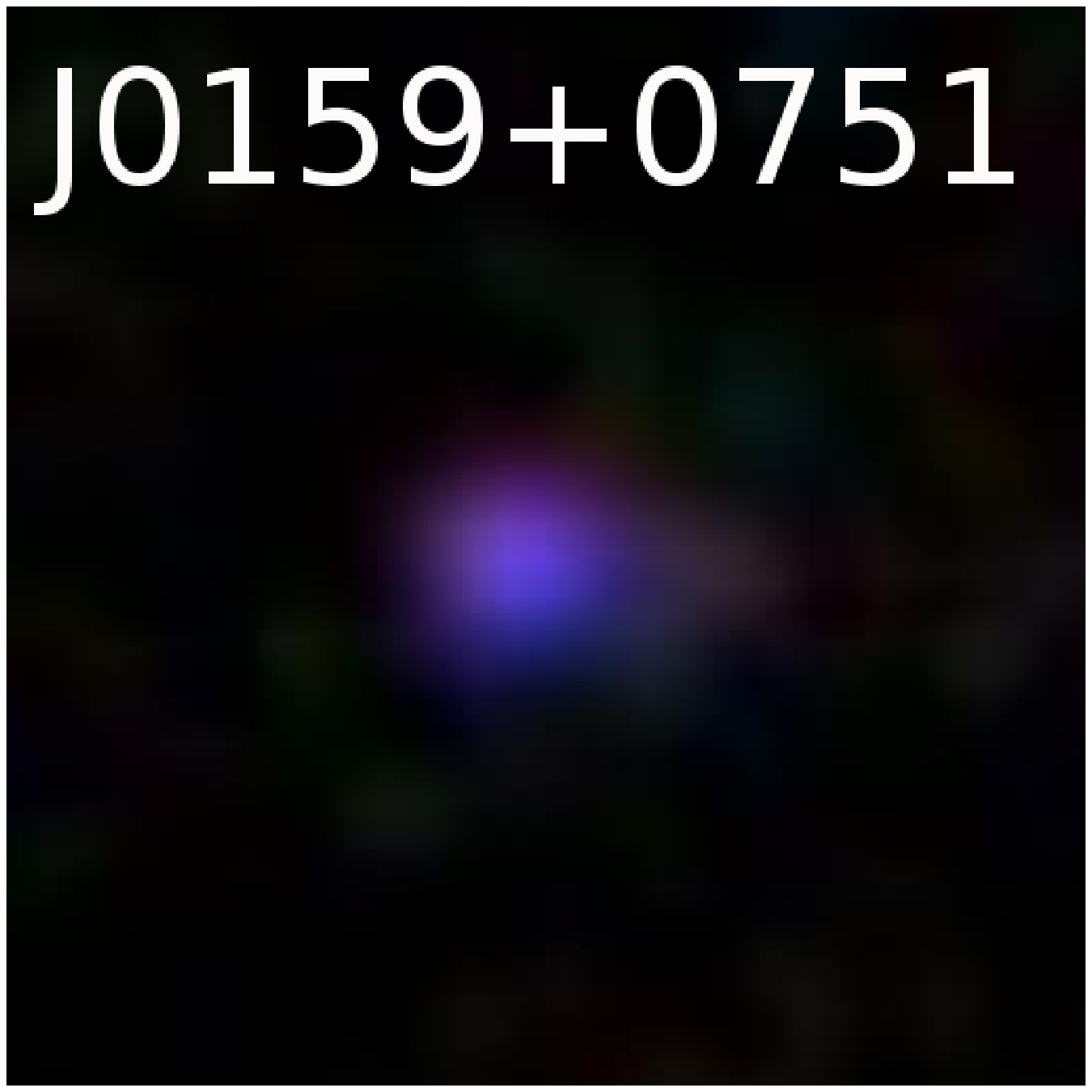}
\includegraphics[angle=0,width=0.19\linewidth]{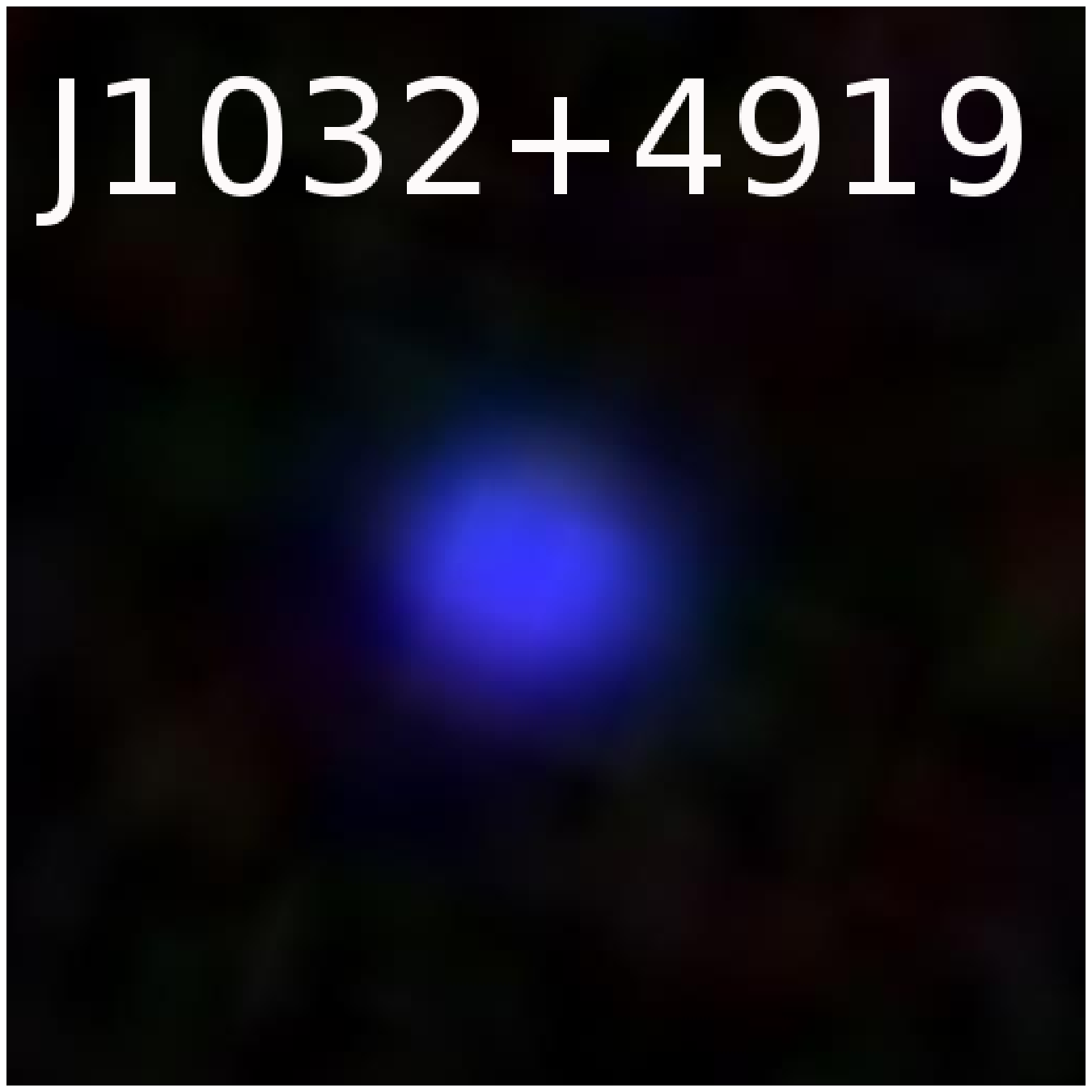}
\includegraphics[angle=0,width=0.19\linewidth]{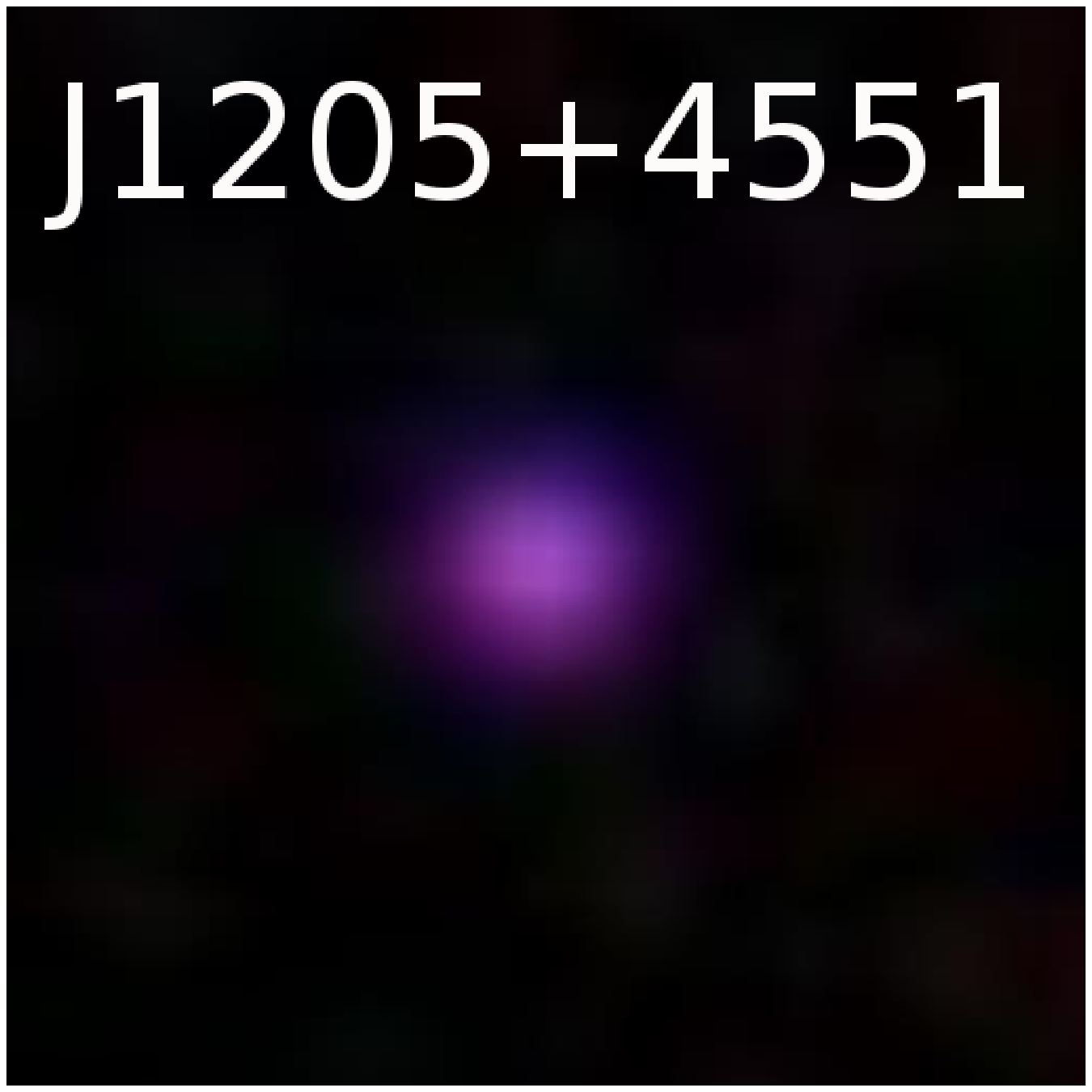}
\includegraphics[angle=0,width=0.19\linewidth]{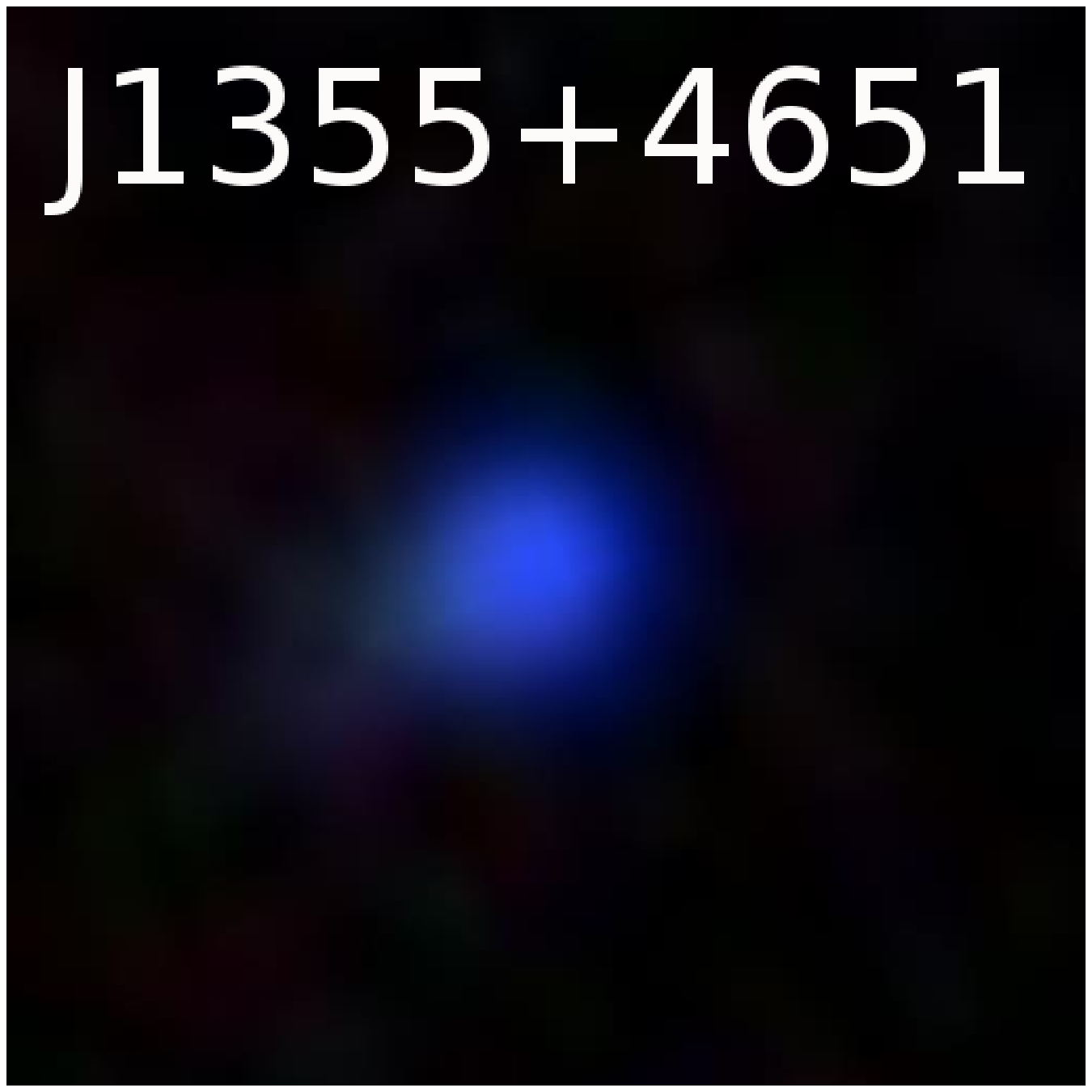}
\includegraphics[angle=0,width=0.19\linewidth]{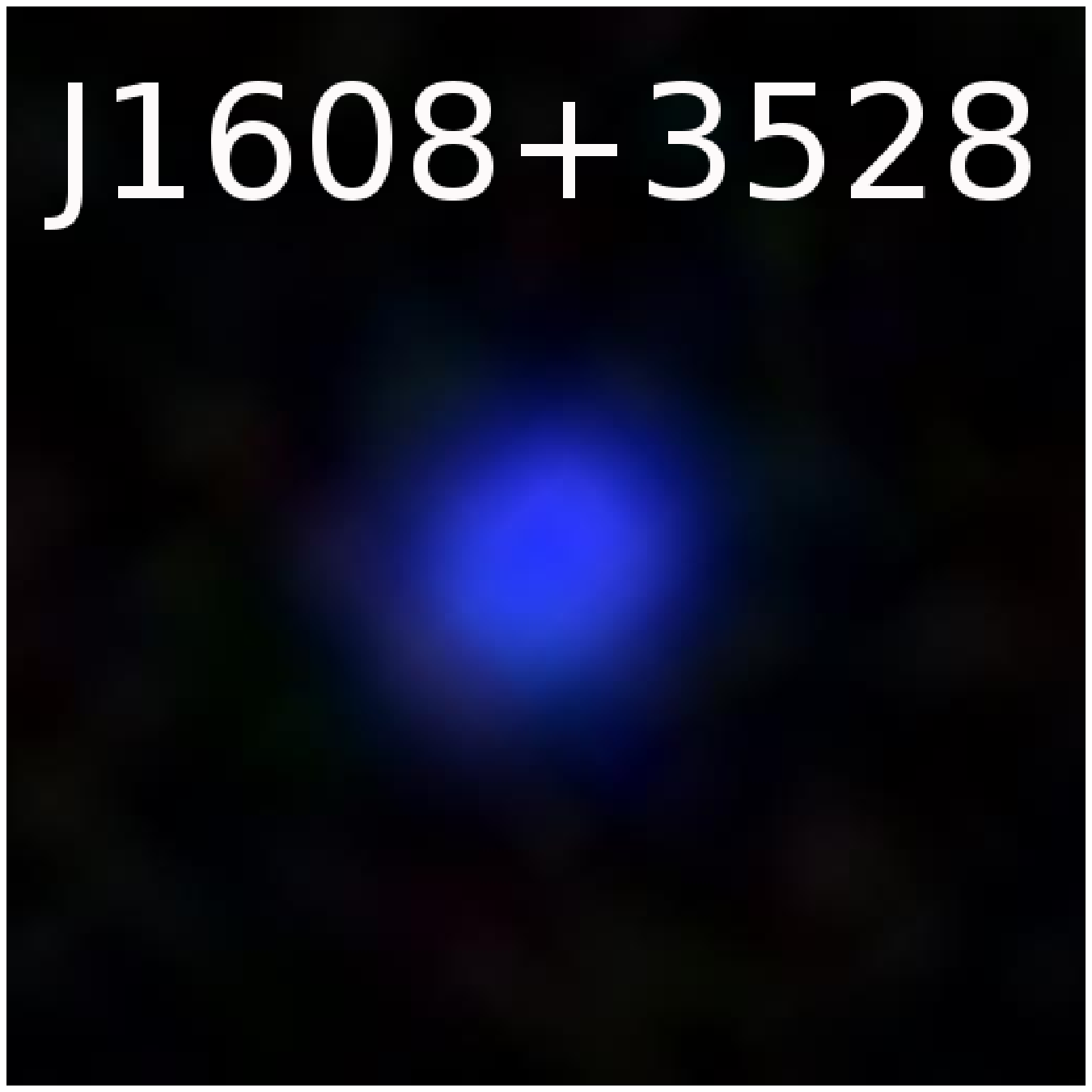}
\caption{The 12\arcsec$\times$12\arcsec\ SDSS images of CSFGs with
extremely high O$_{32}$.}
\label{fig1}
\end{figure*}

\begin{figure*}
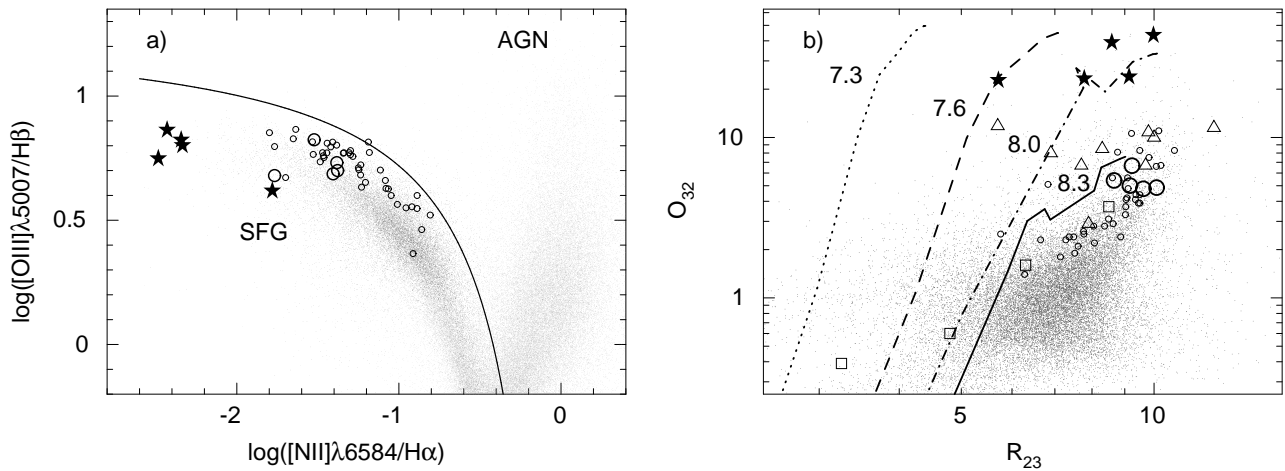

\hbox{
\includegraphics[angle=-90,width=0.47\linewidth]{diagnDR12_1b.ps}
\hspace{0.3cm}\includegraphics[angle=-90,width=0.47\linewidth]{oiii_oii_ca1.ps}
}
\caption{a) The BPT diagnostic diagram
\citep*{BPT81} for narrow emission-line galaxies. CSFGs with extremely 
high O$_{32}$ (this paper) are shown by filled stars.
For comparison, LyC leaking galaxies \citep{I16a,I16b} and confirmed
Ly$\alpha$ leakers \citep{Y17}
are represented by open large and small circles, respectively. Also shown are 
CSFGs from the SDSS DR12
\citep[dark-grey dots, ][]{I16c} and SDSS DR7 narrow emission-line galaxies
(light-grey dots).
The solid line by \citet{K03} separates SFGs from active galactic nuclei
(AGNs). b) The O$_{32}$ - R$_{23}$ diagram for SFGs, 
where R$_{23}$=([O {\sc ii}] 3727 +
[O {\sc iii}]  4959 + [O {\sc iii}] 5007)/H$\beta$. CSFGs with extreme 
O$_{32}$ (this paper) are shown by filled stars. The
location of $z$ $\sim$ 0.3 LyC leaking galaxies \citep{I16a,I16b}
and known low-redshift ($z$ $<$ 0.1) LyC leaking galaxies \citep{L13,Bor14,L16} 
are shown by large open circles and open squares, respectively. 
For comparison, high-$z$ Ly$\alpha$ emitting galaxies (LAEs) by \citet{NO14} 
and \citet{N16} (open triangles), confirmed low-$z$ Ly$\alpha$ emitting galaxies
\citep{Y17} (small open circles) are shown.
SDSS CSFGs \citep{I16c} are shown by grey 
dots. Theoretical predictions of optically-thick (case B) 
photoionized H {\sc ii} region models with log of filling factor of $-$0.5
and with the production rate of ionizing radiation $Q$ = 10$^{53}$ s$^{-1}$ for
four values of the oxygen abundance 12+logO/H = 7.3, 7.6, 8.0, and 8.3 
are shown by dotted, dashed, dash-dotted and solid lines, respectively.}
\label{fig2}
\end{figure*}

\input{tab1_1.tex}

\input{tab2.tex}

\input{tab3.tex}

\begin{figure*}
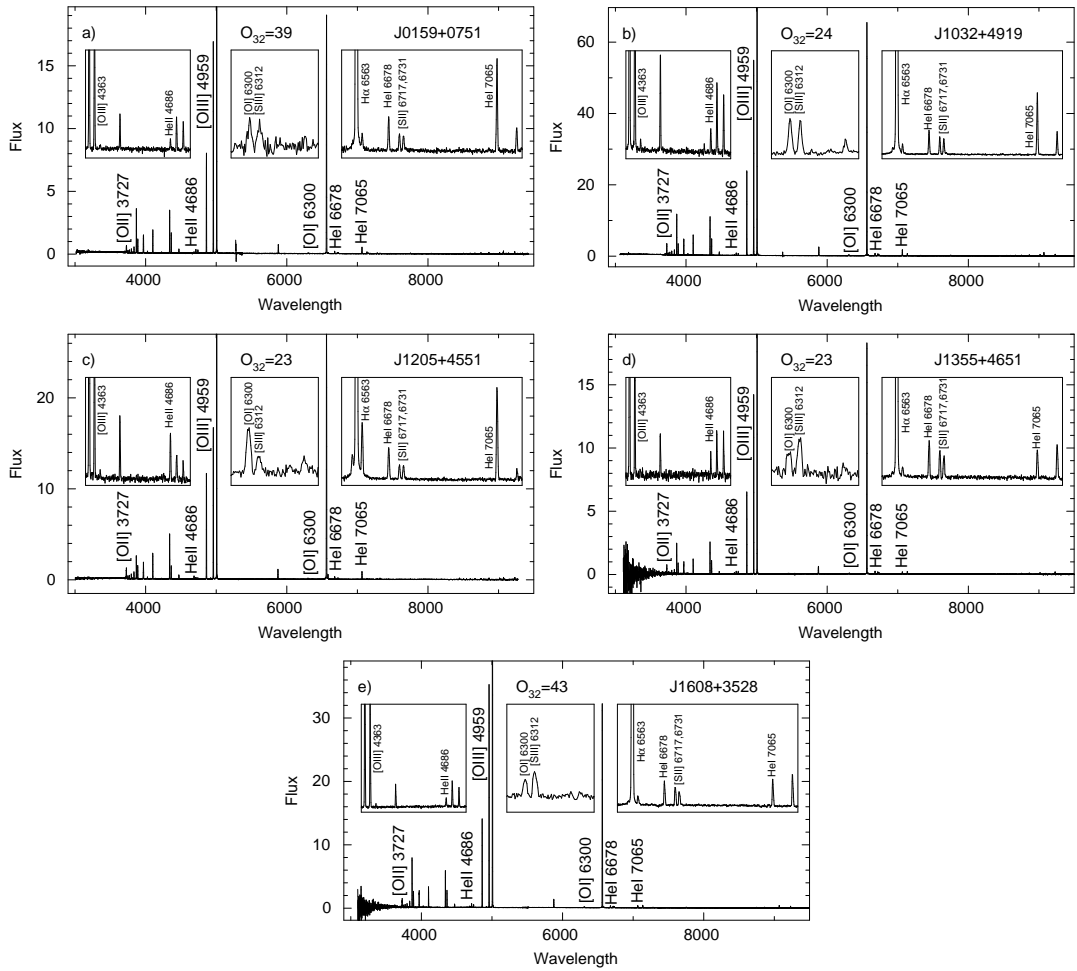

\centering{
\hbox{
\hspace{1.5cm}\includegraphics[angle=-90,width=0.40\linewidth]{spectrum_J0159+0751_1.ps}
\includegraphics[angle=-90,width=0.40\linewidth]{spectrum_J1032+4919_1.ps}}
\hbox{
\hspace{1.5cm}\includegraphics[angle=-90,width=0.40\linewidth]{spectrum_J1205+4551_1.ps}
\includegraphics[angle=-90,width=0.40\linewidth]{spectrum_J1355+4651_1.ps}}
\includegraphics[angle=-90,width=0.40\linewidth]{spectrum_J1608+3528_1.ps}
}
\caption{ The rest-frame LBT spectra of CSFGs with extremely high 
O$_{32}$. Spectra were not corrected for extinction.
Insets show expanded parts of spectral regions of interest in this
paper. From left to the right, they cover the rest 
wavelength ranges 4325\AA\ -- 4770\AA, 6280\AA\ -- 6380\AA, and 
6510\AA\ -- 7155\AA, respectively. Some interesting emission lines are 
labelled. Wavelengths are in \AA\ and fluxes are in 
10$^{-16}$ erg s$^{-1}$ cm$^{-2}$ \AA$^{-1}$.}
\label{fig3}
\end{figure*}

Direct observations of CSFGs at UV wavelengths 
below the Lyman limit were needed to check for the amount of   
escaping Lyman continuum (LyC) radiation. \citet{I16a,I16b} detected escaping 
LyC radiation with escape fraction $f_{\rm esc}$ = 7 -- 13 percent in five 
CSFGs at redshifts $z$ $\sim$ 0.3, using the Cosmic Origin
Spectrograph (COS) aboard the {\sl Hubble Space Telescope} ({\sl HST}). 
Objects need to be in the redshift range of  $\sim$ 0.3 -- 0.4 so as  
to shift the LyC spectral region 
into the wavelength range where the COS is most sensitive.  
These LyC leaking galaxies together with four other known 
LyC leakers at lower redshifts \citep*{L13,Bor14,L16} 
are all characterised by a relatively low O$_{32}$ $\la$ 8. 

Our team has been awarded 
{\sl HST} time in Cycle 24 to observe CSFGs at $z$ $\sim$ 0.3 -- 0.4, with more extreme O$_{32}$, 
$\sim$ 10 -- 30 (Program GO14635, PI: Y.I.Izotov), but the data are not yet 
obtained. The number of such extreme galaxies at $z$ $\ga$ 0.3, suitable for 
{\sl HST} observations, is very limited in the SDSS because the majority 
of them are very faint. Many more brighter extreme SDSS CSFGs can be  
found at lower redshifts, with $z$~$\leq$~0.3. However, their low redshifts 
prevent them from being observed with the {\sl HST} in the Lyman limit
range, because this range is not shifted into the spectral region where the 
COS is most sensitive. Therefore, indirect methods for selecting candidate 
CSFGs with potential LyC leakage are needed. In the UV range, the profile 
shape of the Ly$\alpha$ emission line has been proposed 
as a diagnostic for LyC leaking \citep{V17}. 
However, UV observations of low-redshift objects can only be done from space 
and are not readily available.  
The aim of this paper is to identify line ratio diagnostics
in the optical range which, in addition to the high O$_{32}$ ratio, 
can be used to select potential LyC leakers from the 
SDSS. The best candidates can then be followed-up in the UV   
with the {\sl HST} to study their Ly$\alpha$ emission line profile
and low-ionization (LIS) absorption lines, put constraints
on the optical depth of the LyC and derive the escape fraction of leaking
ionizing radiation.  

In this paper, we have used Large Binocular Telescope (LBT)\footnote{The LBT is an international collaboration among institutions in the United States, Italy and Germany. LBT Corporation partners are: The University of Arizona on behalf of the Arizona university system; Istituto Nazionale di Astrofisica, Italy; LBT Beteiligungsgesellschaft, Germany, representing the Max-Planck Society, the Astrophysical Institute Potsdam, and Heidelberg University; The Ohio State University, and The Research Corporation, on behalf of The University of Notre Dame, University of Minnesota and University of Virginia.}
spectroscopic
observations with high signal-to-noise ratio of five CSFGs selected from the SDSS to have 
extremely high O$_{32}$ = 23 -- 43, to 
develop such optical diagnostics. 
In Section \ref{sec:selection} we describe the sample of CFSGs and the criteria used
for their selection. The observations and data reduction are described in
Section \ref{sec:observations}. Element abundances are derived in Section
\ref{sec:abundances}. In Section \ref{sec:broad} we study the H$\alpha$ 
broad-line emission. The emission-line diagnostics, which can be used to  
identify potential LyC leakers, are discussed in Section 
\ref{sec:diagnostics}. We summarize our main results in Section 
\ref{sec:conclusions}.

\input{tab4_1.tex}

\section{Selection of CSFGs and their global characteristics}\label{sec:selection}

From the 
spectroscopic data base of the SDSS Data Release 12 
(DR12) \citep{A15}, \citet{I16c} have 
constructed a sample of $\sim$ 15000 CSFGs.
 These objects were selected by their compactness, strong 
emission lines and absence of AGN spectral features. 
The sample includes 26 galaxies
with extremely high O$_{32}$, $>$ 20. Out of these, five galaxies have been 
chosen for LBT observations. Their coordinates, O$_{32}$, redshifts and apparent
SDSS $g$ magnitudes are presented in Table \ref{tab1}. 
The composite SDSS images of the selected galaxies are shown in Fig. \ref{fig1}.
They are all characterised by a compact nearly unresolved structure.
All five galaxies
were detected by {\sl Galaxy Evolution Explorer} ({\sl GALEX}) in the UV range 
and by {\sl Wide-field Infrared Survey Explorer} ({\sl WISE}) in the 
mid-infrared (MIR) range. Their UV and MIR apparent 
magnitudes are shown in Table \ref{tab1}. Out of the five CSFGs, three objects, 
J0159+0751, J1032+4919 and J1205+4551, have very red
 {\sl WISE} colours, $W1-W2$ $>$ 1.5 mag, indicating the presence of 
hot dust \citep{I14a}. The $W1-W2$ colours of the remaining two CSFGs,
J1355+4651 and J1608+3528, are however not so red ($<$ 1 mag), implying less
dust heating.

The location of the five selected galaxies in the 
[O~{\sc iii}]$\lambda$5007/H$\beta$ -- [N~{\sc ii}]$\lambda$6584/H$\alpha$ 
diagnostic diagram \citep*{BPT81} is shown in Fig. \ref{fig2}a by
filled stars. It is seen that four out of the five CSFGs are more 
extreme objects compared to the five strong LyC leakers from \citet{I16a,I16b} (large open circles), to the confirmed Ly$\alpha$ emitting
GPs of \citet{Y17} (small open circles) and to the entire sample of CSFGs 
\citep[dark-grey dots, ][]{I16c}. Only one galaxy, J1205+4551, with a higher 
[N~{\sc ii}]$\lambda$6584/H$\alpha$ ratio and a lower 
[O~{\sc iii}]$\lambda$5007/H$\beta$ ratio, has properties similar to 
the most extreme LyC leakers and Ly$\alpha$ emitting GPs. The solid line 
derived by \citet{K03} separates SFGs from active galactic nuclei (AGN). 
All five CSFGs are located in the SFG region, implying that  
their interstellar medium is ionized by hot stars in the star-forming regions. 
However, they lie somewhat below the main sequence defined by the other 
SFGs in Fig.~\ref{fig2}a, a consequence of their lower metallicities.

In Fig. \ref{fig2}b we compare the locations in the 
O$_{32}$ -- R$_{23}$ 
(R$_{23}$=\{[O~{\sc ii}]$\lambda$3727 + [O~{\sc iii}]$\lambda$4959 + 
[O~{\sc iii}]$\lambda$5007\}/H$\beta$) diagram of CSFGs possessing  
extremely high O$_{32}$ (filled stars) with those of 
high-redshift LAEs that are potential LyC leakers
\citep{NO14,N16} (open triangles), of known $z$ $\sim$ 0.3 LyC leakers
\citep[large open circles, ][]{I16a,I16b}, of LyC leaking galaxies at 
$z$ $<$ 0.1 \citep[open squares, ][]{L13,Bor14,L16}, and of Ly$\alpha$
emitting GPs \citep[small open circles, ][]{Y17}. We show by dotted, dashed, 
dash-dotted and solid lines the evolutionary sequences predicted by
{\sc Cloudy} photoionization models calculated for four values of the oxygen
abundance (the 12+log O/H value labels each curve),
for H~{\sc ii} regions ionized only by stellar radiation
and that are optically thick (case B). Along the curves, higher modelled 
O$_{32}$ values correspond to younger starburst ages and thus to more
intense and harder ionizing radiation.
It is seen that the selected galaxies are located far 
above the LAEs, at the extreme of the distribution for SDSS galaxies, implying 
that they may be good LyC leaking candidates. 
In Fig. \ref{fig2}b, they lie somewhat apart from the main sequence 
defined by the other CSFGs because of their lower metallicities.
No galaxy with such extreme O$_{32}$ has been studied until now.

The global characteristics of the CSFGs with extreme O$_{32}$, derived from
their SDSS spectra and {\sl GALEX} photometry, are shown in Table \ref{tab2}.
The observed fluxes have been transformed to luminosities and absolute
magnitudes, adopting luminosity distances \citep[NED,][]{W06} derived with
the cosmological parameters $H_0$=67.1 km s$^{-1}$Mpc$^{-1}$, 
$\Omega_\Lambda$=0.682, $\Omega_m$=0.318 \citep{P14}. 
The H$\beta$ luminosities $L$(H$\beta$) were derived from the 
extinction-corrected H$\beta$ fluxes measured in the SDSS spectra. Additionally,
$L$(H$\beta$) were also corrected for aperture effects using the relation 
2.512$^{r({\rm ap})-r}$, where $r$ and $r$(ap) are respectively the SDSS 
$r$-band total magnitude and the magnitude 
within the round spectroscopic SDSS 3 arcsec aperture
for galaxies in the SDSS-II and 2 arcsec aperture for galaxies
in the SDSS-III.
We have also derived extinction-corrected 
absolute AB SDSS $g$-band and {\sl GALEX} FUV magnitudes.

\begin{figure*}
\centering{
\includegraphics[angle=-90,width=0.99\linewidth]{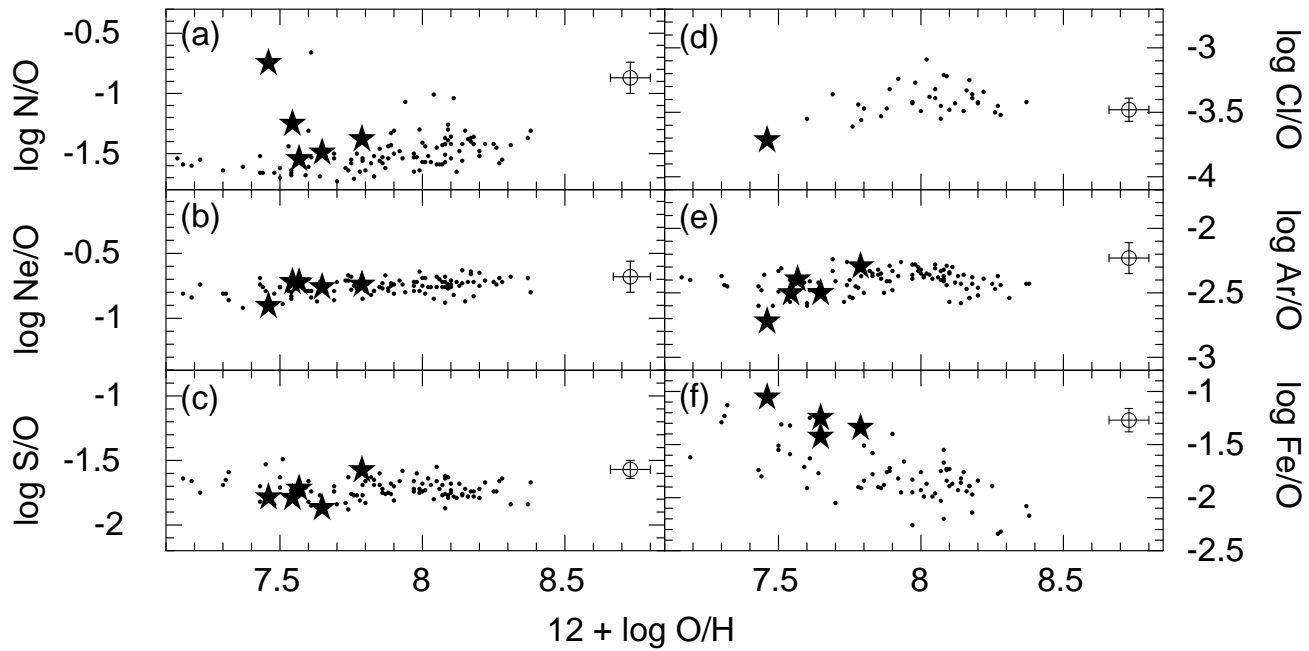}
}
\caption{The dependences of heavy element abundance ratios (a) N/O, (b) Ne/O,
(c) S/O, (d) Cl/O, (e) Ar/O and (f) Fe/O on the oxygen
abundance 12+logO/H. CSFGs (this paper) are shown by filled stars.
For comparison, BCDs from \citet{I06} and solar values by \citet{L10}
are shown by dots and open circles, respectively.}
\label{fig4}
\end{figure*}

Using a Monte Carlo method, we have calculated galaxy stellar masses by 
fitting the spectral energy distribution (SED) of the SDSS spectra 
corrected for extinction as derived from the observed hydrogen Balmer decrement
from the same spectra. The spectral fluxes
 were also corrected for the finite size of the 
spectroscopic aperture by 
comparing the total SDSS $r$ magnitude to the magnitude
$r_{\rm ap}$ inside the aperture.
The Monte Carlo method is described e.g. in \citet{I11,I14a}. Briefly, 
a grid of instantaneous burst SEDs in a wide range of ages, 
ranging from 0 Myr to 
15 Gyr, was calculated with the {\em Starburst99} code \citep{L99,L14} to derive
the SED of the galaxy stellar component. We adopted Padova stellar 
evolution tracks \citep{G00}, stellar atmosphere models by \citet*{L97},
and the stellar initial mass function by \citet{S55}.
Then the stellar SED with any star-formation 
history can be obtained by integrating the instantaneous burst SEDs over
time with a specified time-varying star formation rate. 
Given the electron temperature $T_{\rm e}$ in the H~{\sc ii} region, we 
interpolated emissivities from \citet{A84} for the nebular continuum 
in the $T_{\rm e}$ range of 5000 -- 20000K, including hydrogen 
and helium free-bound, free-free, and two-photon emission. The emission lines
were added to the total, stellar and nebular, continuum with fluxes measured 
in the SDSS spectra.
The star-formation history is approximated assuming a
short burst with age $<$ 10 Myr and a prior continuous star formation
with a constant SFR.

It is seen from Table \ref{tab2} that the selected galaxies are dwarf systems with
low stellar masses $M_\star$, ranging from $\sim$~10$^6$ to $\sim$~10$^7$ 
M$_\odot$. These stellar masses are $\sim$ 2 orders of magnitude 
lower than the stellar masses 
of the $z$ $\sim$ 0.3 LyC emitters studied by \citet{I16a,I16b}. 
Furthermore, their
absolute magnitudes are $\sim$ 2 mag fainter.
The extreme CSFGs are 
much more metal-poor, with oxygen abundances 12+logO/H between 7.46 and 
7.79 (see also section \ref{sec:abundances}).
Finally, the equivalent widths of the H$\beta$ emission line in our five
CSFGs are very high, $\sim$ 350 -- 520\AA\ (see section \ref{sec:observations})
implying a very high
efficiency of the ionizing photon production $\xi$. These properties are 
similar to those of some local BCDs \citep{T08} and of 
the hypothetical dwarf galaxies thought to be the 
main contributors to the reionization of the Universe at redshifts $z$ $>$ 5.

\input{tab5.tex}

\section{Observations and data reduction}\label{sec:observations}

In the course of four LBT runs in the period 2013 -- 2016, we have obtained
long-slit spectrophotometric observations in the wavelength range
3200 -- 10000\AA\ of the galaxies listed in Table \ref{tab1} with
the dual spectrograph MODS1\footnote{This paper used data obtained with the MODS spectrographs built with
funding from NSF grant AST-9987045 and the NSF Telescope System
Instrumentation Program (TSIP), with additional funds from the Ohio
Board of Regents and the Ohio State University Office of Research.}, equipped 
with two 8022 pix $\times$ 3088 pix CCDs. The G400L grating in the 
blue beam, with a dispersion of 0.5\AA/pix, and the G670L grating in the 
red beam, 
with a dispersion of 0.8\AA/pix, were used.
Spectra were obtained with a 1.2 arcsec wide slit, 
resulting in a resolving power $R$ $\sim$ 2000.

Three or four 800~s subexposures were obtained for each galaxy, resulting
in total exposures of 2400~s or 3200~s (Table~\ref{tab3}). 
The airmass  was small during the observations of three CSFGs (Table \ref{tab3}). However, it is somewhat larger ($\sim$ 1.4) during the 
observations of J1355+4651 and J1608+3528, so that 
their spectra may be affected by atmospheric refraction. Several 
spectrophotometric standard stars, obtained during the same nights
with a 5 arcsec wide slit, were used for flux calibration and 
correction for telluric absorption lines in the red part of the spectra. 
Additionally, calibration frames of biases, flats and argon comparison lamps 
were obtained during the same nights.

Basic data reduction was done with the MODS Basic CCD Reduction package
{\sc modsCCDRed}\footnote{http://www.astronomy.ohio-state.edu/MODS/Software/~modsCCDRed/}, 
including bias subtraction and flat field correction. 
Subsequent data reduction including wavelength and flux calibration, 
and night sky background subtraction was
done with {\sc IRAF}\footnote{{\sc IRAF} is distributed by the 
National Optical Astronomy Observatories, which are operated by the Association
of Universities for Research in Astronomy, Inc., under cooperative agreement 
with the National Science Foundation.}. Cosmic ray hits were manually
removed from the background-subtracted frames which were then flux-calibrated. 
Individual subexposures for each object were co-added. Finally,
one-dimensional spectra were extracted using the {\sc IRAF} {\sc apall} routine.

The resulting rest-frame spectra are shown in Fig.~\ref{fig3}. They are
dominated by strong emission lines, suggesting active star formation.
The strong [O {\sc iii}] $\lambda$4363 emission line is present in all spectra,
allowing reliable abundance determination. Each panel includes three insets
showing expanded parts of the spectra for a better view of particular 
emission lines of interest. There are several features of note. 
First, a weak broad component of the H$\alpha$ emission line is
seen in the spectra of all CSFGs. Second, the strength of
the He {\sc i} $\lambda$7065 emission line relative to the 
He {\sc i} $\lambda$6678 emission line
varies in a large range. The former line depends more on collisional
and fluorescent enhancement than the latter, 
and variations in their intensity ratios indicate
 that the physical conditions in the H {\sc ii}
regions of the studied galaxies, 
as characterized by the electron number density and the optical depth, 
vary over a wide range. 
Third, an important
feature of all spectra is the 
very weak [O {\sc ii}] $\lambda$3727 emission line, as compared to the 
strong 
[O {\sc iii}] $\lambda$4959, $\lambda$5007 emission lines,
resulting in unusually high O$_{32}$. Fourth, in one of the CSFGs, J1205$+$455,
the high-ionization emission line [Ne {\sc v}] $\lambda$3426\AA\ is detected,
implying the presence of shocks produced by supernovae \citep{I12}.

Emission-line fluxes were measured using the {\sc IRAF} {\sc splot} routine. 
The errors of the line fluxes were calculated from the
photon statistics in the non-flux-calibrated spectra and adding a 
relative error of 1 percent in the 
absolute flux distribution of the spectrophotometric standards.
The line flux errors were propagated in the calculations of the elemental 
abundance errors. 

The observed fluxes were corrected for extinction with the extinction 
coefficient $C$(H$\beta$),  derived from the observed decrement of the
hydrogen Balmer emission lines.
We have adopted the \citet*{C89} reddening law with 
$R(V)$ = $A(V)$/$E(B-V)$ = 3.1, where $A(V)$ and $E(B-V)$ are respectively the 
total extinction in the $V$ band and the selective extinction.
Recently, \citet{I17} by analyzing UV data of CSFGs, have shown 
that $R(V)$ may be lower than 3.1 in the UV range. However, in the optical 
range, at variance with the UV range, reddening laws with different $R(V)$'s 
are quite similar. For $R(V)$ = 3.1, the total extinction is linked to 
$C$(H$\beta$) by the
relation $A(V)$ = 2.11$\times$$C$(H$\beta$). The relations for other
$R(V)$'s are given in \citet{I16a}. We have also assumed that extinctions
for nebular and stellar emission are equal, in accord with \citet{I17}.
The extinction-corrected emission line fluxes $I$($\lambda$)
relative to the H$\beta$ fluxes multiplied by 100, the extinction coefficients
$C$(H$\beta$), the rest-frame equivalent widths EW(H$\beta$) and the observed
H$\beta$ fluxes $F$(H$\beta$) are listed in Table \ref{tab4}. We note that 
the EW(H$\beta$)s of the 
studied CSFGs are very high, $\sim$~350 -- 520\AA. They are 
among the highest ever measured in spectra of SFGs, indicating
the bursting nature of the star formation and very young burst ages ($<$ 3 Myr).
For comparison, rest-frame EW(H$\beta$) of $\sim$ 200\AA\ were found by 
\citet{I16a,I16b} for the five $z$ $\sim$ 0.3 LyC leakers, indicating 
somewhat larger burst ages.

We note, however, that the LBT spectra of J1355$+$4651 J1608$+$3528 
obtained at high air masses are affected by atmospheric refraction. Indeed, the
extinction coefficients $C$(H$\beta$) for these two objects derived from the 
SDSS spectra with a wider 3 arcsec aperture are somewhat lower, being 
respectively 0.065 and 0.20. The respective values of O$_{32}$ from the SDSS 
spectra are also lower, being 21 and 35. The effect of the atmospheric 
refraction is especially important in the LBT spectrum of J1355$+$4651, 
therefore its high $C$(H$\beta$) of 0.65 is likely overestimated. Taking 
into account this refraction effect, we conclude that the overall
dust extinction in the galaxies studied with the LBT is low.

\section{Element abundances}\label{sec:abundances}

To determine element abundances, we follow the procedures of \citet{I06}.
We adopt a two-zone photoionized H {\sc ii} region model: a high-
ionization zone with temperature $T_{\rm e}$(O {\sc iii}), where [O {\sc iii}],
[Ne {\sc iii}], and [Ar {\sc iv}] lines originate, and a low-ionization zone 
with temperature $T_{\rm e}$(O {\sc ii}), where [N {\sc ii}], 
[O {\sc ii}], [S {\sc ii}],
and [Fe {\sc iii}] lines originate. As for the [S~{\sc iii}] and [Ar {\sc iii}]
lines, they originate in the intermediate zone between the
high- and low-ionization regions. The temperature $T_{\rm e}$(O {\sc iii})
is calculated using the 
[O~{\sc iii}] $\lambda$4363/($\lambda$4959 + $\lambda$5007) ratio.
To take into account the electron temperatures for different
ions, we have used the expressions of \citet{I06}.
The electron number density $N_{\rm e}$(S {\sc ii}) is derived from the 
[S {\sc ii}] $\lambda$6717/$\lambda$6731 emission line ratio.
However, we were not able to derive the electron number density from
the [O {\sc ii}]$\lambda$3726,3729 doublet because of insufficient spectral
resolution.

The electron temperatures $T_{\rm e}$(O {\sc iii}) in the studied CSFGs are high
(Table \ref{tab5}),
ranging from $\sim$17000K to $\sim$21000K. The electron number
density $N_{\rm e}$(S {\sc ii}), characteristic of the low-ionization
zone of the H {\sc ii} region, is also relatively high. It ranges from 
470 cm$^{-3}$ to 640 cm$^{-3}$ in J0159+0751, J1032+4919 and J1205+4551 and is
somewhat lower, 190 cm$^{-3}$ and 260 cm$^{-3}$ in J1355+4651 and 
J1608+3528, respectively. This is to be compared 
with average values of $\leq$~100 cm$^{-3}$ generally found  
for H {\sc ii} regions in SFGs.
Thus, the H {\sc ii} regions in the selected galaxies with extreme O$_{32}$
are hot and dense. 

Ionic and total heavy element abundances are presented in Table \ref{tab5}, 
together with the ionization
correction factors $ICF$ for unseen stages of ionization. The derived oxygen
abundances are low, ranging from 12+logO/H = 7.46 to 7.79. These values are
lower than those derived by 
\citet{I16a,I16b} for the $z$ $\sim$ 0.3 LyC leaking 
galaxies (between 7.8 and 8.0). The other heavy element to oxygen abundance ratios 
are shown in Fig. \ref{fig4}. Except for nitrogen 
(Fig. \ref{fig4}a), the abundance ratios 
for the CSFGs studied in this paper (filled stars) are similar to those 
derived for a sample of BCDs with high signal-to-noise
ratio spectra \citep{I06} shown by dots in Fig. \ref{fig4}. 
We note however that the
Ar/O ratio for the galaxy J1205+4551 (the most metal-poor object in 
Fig.~\ref{fig4}e) may somewhat be underestimated. This is because the Ar
abundance is determined from the [Ar~{\sc iii}] $\lambda$7135\AA\ emission line,
which is redshifted to the wavelength $\sim$ $\lambda$7600\AA\ where the 
telluric absorption is strong. Although we corrected the galaxy 
spectrum for this effect using a standard star spectrum, one should keep in mind that the 
spectrum of the standard star has a worse spectral resolution, 
having been obtained with a 5 arcsec wide slit, while the galaxy was observed with a considerably narrower 1.2 arcsec wide slit.

\begin{figure*}
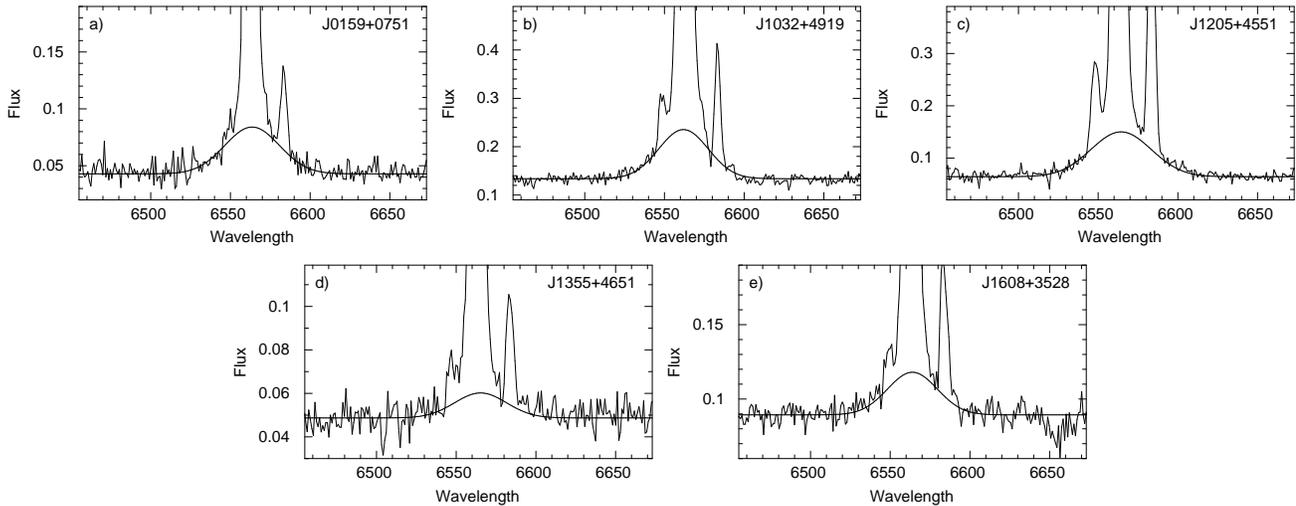

\centering{
\hbox{
\includegraphics[angle=-90,width=0.32\linewidth]{Ha_J0159+0751.ps}
\includegraphics[angle=-90,width=0.32\linewidth]{Ha_J1032+4919.ps}
\includegraphics[angle=-90,width=0.32\linewidth]{Ha_J1205+4551.ps}}
\hbox{
\hspace{3.0cm}\includegraphics[angle=-90,width=0.32\linewidth]{Ha_J1355+4651.ps}
\includegraphics[angle=-90,width=0.32\linewidth]{Ha_J1608+3528.ps}}
}
\caption{ The observed H$\alpha$ profiles with fits of their broad
components.}
\label{fig5}
\end{figure*}

\input{tab6.tex}

The only element which shows a large dispersion is nitrogen
(Fig. \ref{fig4}a). While the three highest-metallicity objects in our
sample have N/O abundance ratios similar to those obtained for 
the BCD sample, a higher N/O abundance ratio is found for 
the two lowest-metallicity 
galaxies. The most extreme case is the galaxy J1205+4551, with a N/O ratio
$\sim$ 0.8 dex higher than the average value for BCDs at that metallicity.
Such a large difference cannot be explained by the nitrogen enrichment of an 
uniform H {\sc ii} region, with a constant electron number density \citep{I06}.
This is however possible in inhomogeneous H {\sc ii} regions with, for example, 
dense clumps enriched in nitrogen around massive stars
with stellar winds. In this case, nitrogen emission lines would be enhanced
because of both higher nitrogen abundances and higher electron number densities
in clumps, while oxygen lines would only be 
enhanced by the latter. Therefore, in such H {\sc ii} regions, the
enhanced N/O abundance ratio 
is not characteristic of the entire H {\sc ii} region, but mostly of 
nitrogen-enriched clumps.

\section{Broad H$\alpha$ line emission}\label{sec:broad}

One of the most interesting features of the CSFG spectra with 
extreme O$_{32}$ is the
presence of a low-intensity broad component of the H$\alpha$ emission line 
in all of them 
(see insets in Fig. \ref{fig3}). These broad components with intensities 
$\sim$ 1 -- 10 percent that of the H$\alpha$ line  
are seen in a minority of SFG spectra \citep*{I07}.
They imply the presence in these galaxies of  
rapid dynamical processes, either expanding envelopes of massive stars
with stellar winds or supernova remnants in their early
stage of expansion.   
Our spectra do not show the presence of broad
blue ($\sim$ $\lambda$4650\AA) or red ($\sim$ $\lambda$5800\AA) bumps, 
characteristic of Wolf-Rayet stars (see insets in Fig. \ref{fig3}),
thus ruling out the hypothesis of massive stars with stellar winds. 
 Therefore, the most likely mechanism 
for broad  H$\alpha$ emission appears to be  
expanding young supernova remnants in the early stages of their evolution.

In Fig. \ref{fig5} we present for all five CSFGs 
the spectral region around the H$\alpha$ emission
line, with the Gaussian fit to the broad component. The flux
ratios of broad-to-total H$\alpha$ emission and the full widths at half maximum (FWHM) in km s$^{-1}$ of the broad emission are shown in 
Table \ref{tab6}. We see that 
the fraction of broad emission is small ($<$ 3 percent), but that  
expansion velocities are high, with FWHMs ranging from $\sim$1700 to $\sim$2000 km s$^{-1}$.
Such rapid ionized gas motions may facilitate escape of the resonant Ly$\alpha$ emission from the galaxy. They should be taken into account
in models of its transfer through the H {\sc ii} region.

\input{tab7.tex}

\begin{figure*}
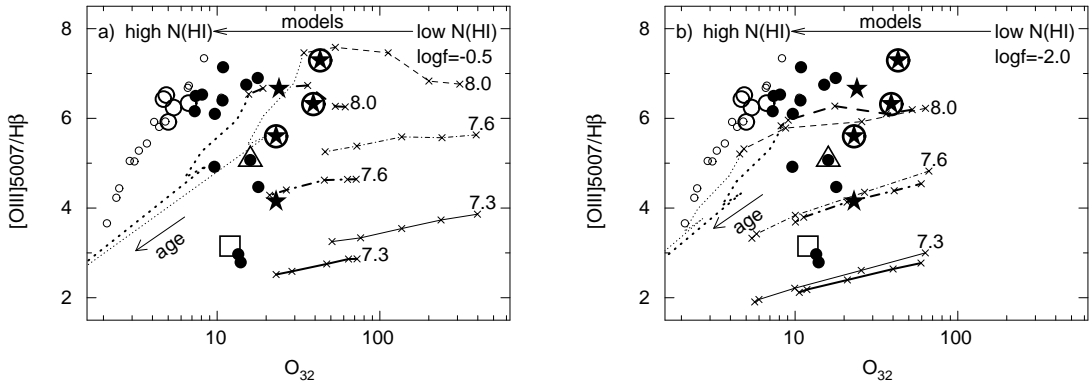

\centering
\includegraphics[angle=-90,width=0.40\linewidth]{o32_O3_1a.ps}
\hspace{0.5cm}\includegraphics[angle=-90,width=0.40\linewidth]{o32_O3_1b.ps}
\caption{The [O~{\sc iii}]5007/H$\beta$ -- O$_{32}$
diagram for a) $\log f$ = $-$0.5 and b) $\log f$ = $-$2.0. 
The five CSFGs with extremely high O$_{32}$ studied here are 
shown by filled stars. The encircled filled stars correspond to CFSGs with small neutral
hydrogen column densities
$N$(H~{\sc i}), as predicted from He {\sc i} line ratios, while the non-encircled stars correspond to those with high $N$(H~{\sc i}). The 
five low-$z$ LyC leakers studied by \citet{I16a,I16b} are 
shown by large open circles. The BCDs 
SBS~0335$-$052 with a Ly$\alpha$ line in absorption, and Tol~1214$-$277 with
a prominent Ly$\alpha$ emission line \citep{TI97}, are shown by an open
square and an open triangle, respectively. A sample of BCDs with O$_{32}$ $\sim$
5 -- 20, observed with the LBT (Izotov et al., 2017, in preparation)
is represented by filled circles, while confirmed Ly$\alpha$ emitting galaxies
\citep{Y17} are shown by small open circles.
{\sc Cloudy} models for a burst with the age of 2 Myr and fixed 
ionization parameters
with three values of the oxygen abundance 12+logO/H = 7.3, 7.6
and 8.0 are shown by solid, dash-dotted and dashed lines, respectively, and 
those with two values of the electron number density $N_{\rm e}$ = 
100 and 1000 cm$^{-3}$ are represented by thin and thick lines, respectively.
Crosses on the modelled lines correspond to 
$N$(H~{\sc i}) ranging from 10$^{17}$ cm$^{-2}$ to 10$^{19}$ cm$^{-2}$, in steps of
0.5 dex and increasing from right to left. The dotted lines show
the evolutionary sequences calculated for {\sc Cloudy}  
optically thick models adopting $N_{\rm e}$ = 1000 cm$^{-3}$ (thick dotted line)
and $N_{\rm e}$ = 100 cm$^{-3}$ (thin dotted line) with 12+logO/H = 8.0 and 
starburst ages varying from 1 Myr to 6 Myr. 
The age direction is given by the arrow.}
\label{fig6}
\end{figure*}

\begin{figure*}
\centering
\includegraphics[angle=-90,width=0.40\linewidth]{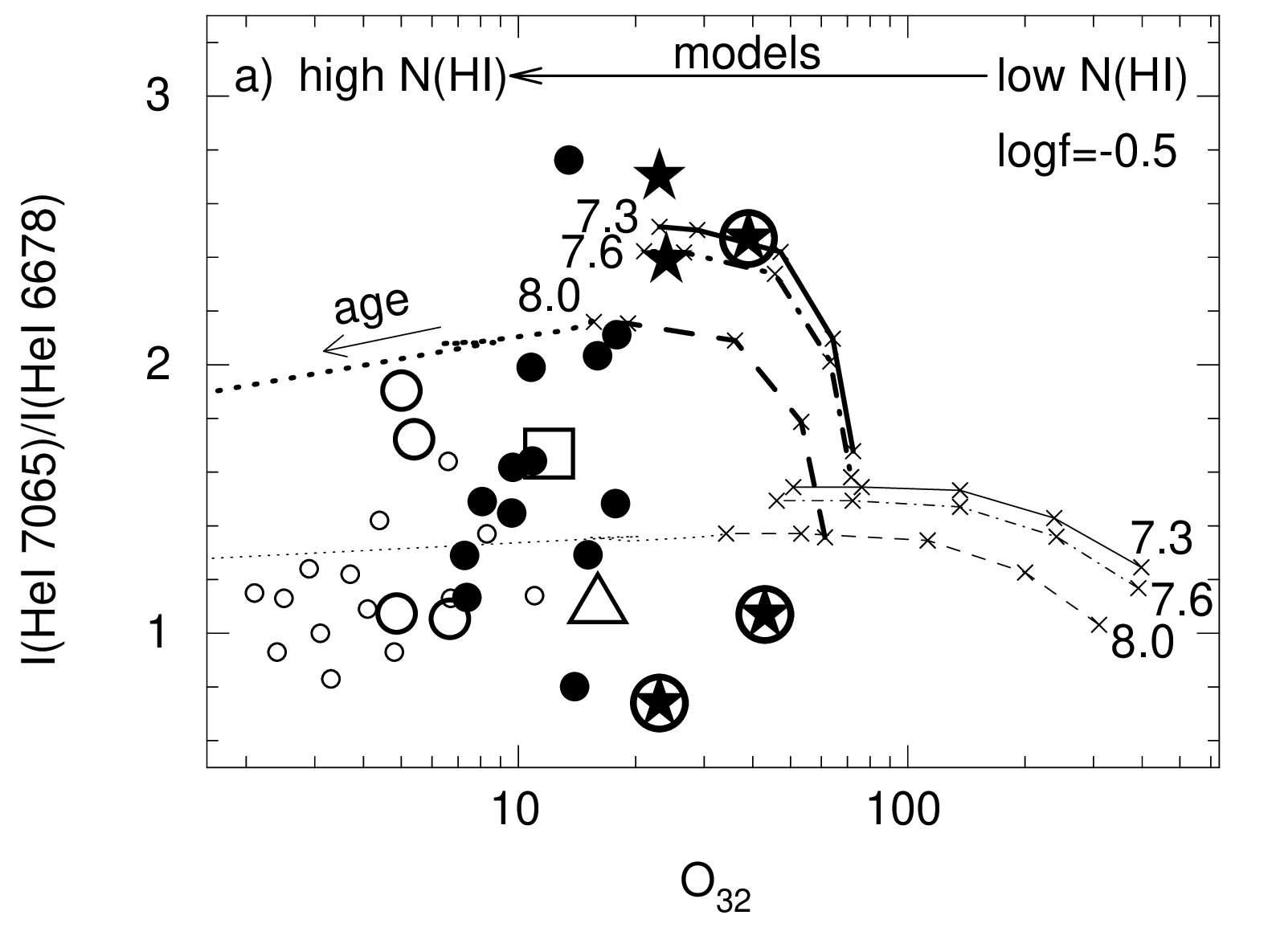}
\hspace{0.5cm}\includegraphics[angle=-90,width=0.40\linewidth]{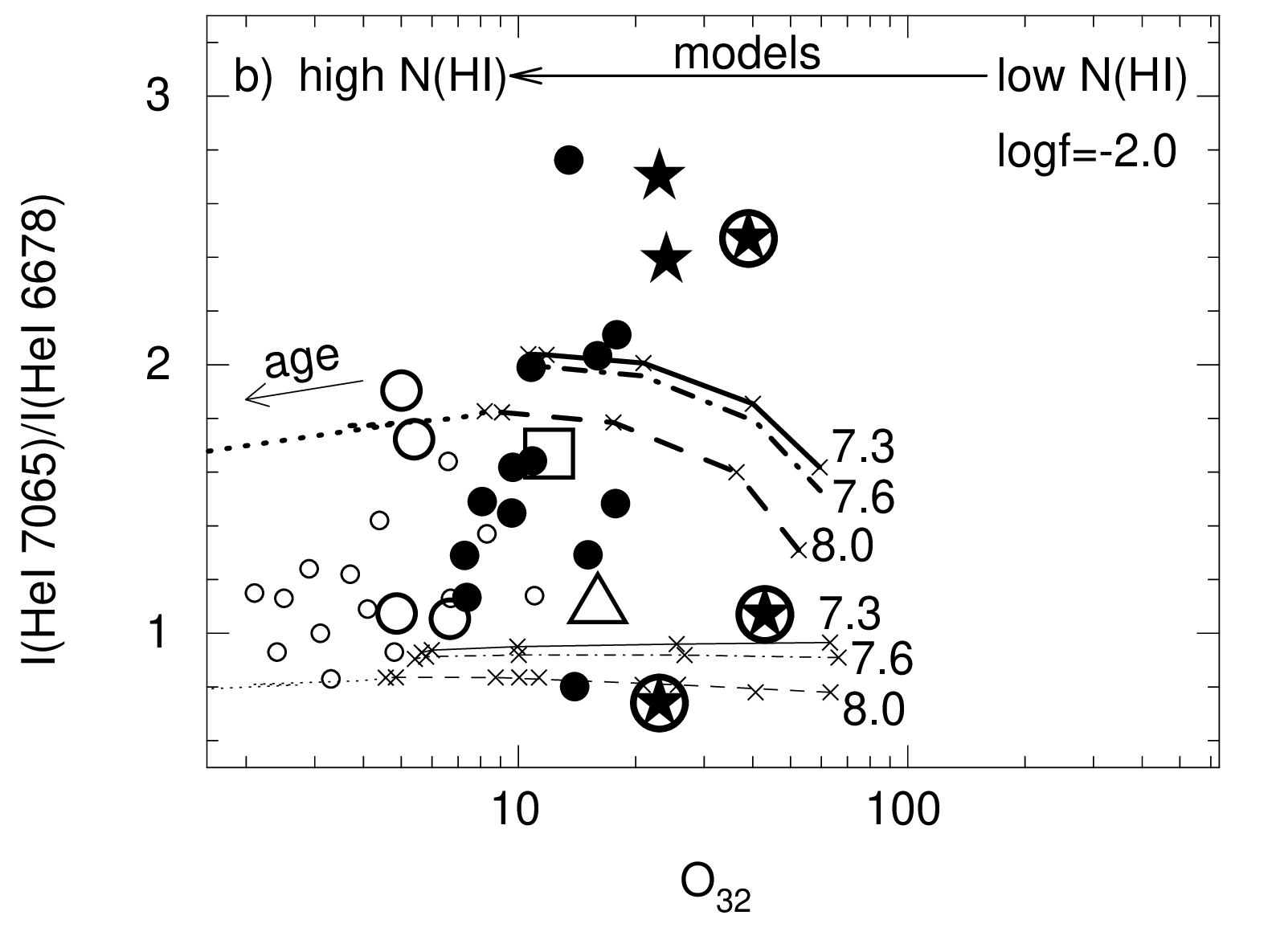}
\caption{The $I$(He {\sc i} $\lambda$7065)/$I$(He {\sc i} $\lambda$6678) 
-- O$_{32}$ diagram for a) $\log f$ = $-$0.5 and b) $\log f$ = 
$-$2.0. Symbols and lines are the same as in Fig.~\ref{fig6}.
The LyC leaker J1333+6246 with noisy He {\sc i} $\lambda$6678 and 
$\lambda$7065 lines in its SDSS spectrum is not shown.
}
\label{fig7}
\end{figure*}

\begin{figure*}
\centering
\includegraphics[angle=-90,width=0.40\linewidth]{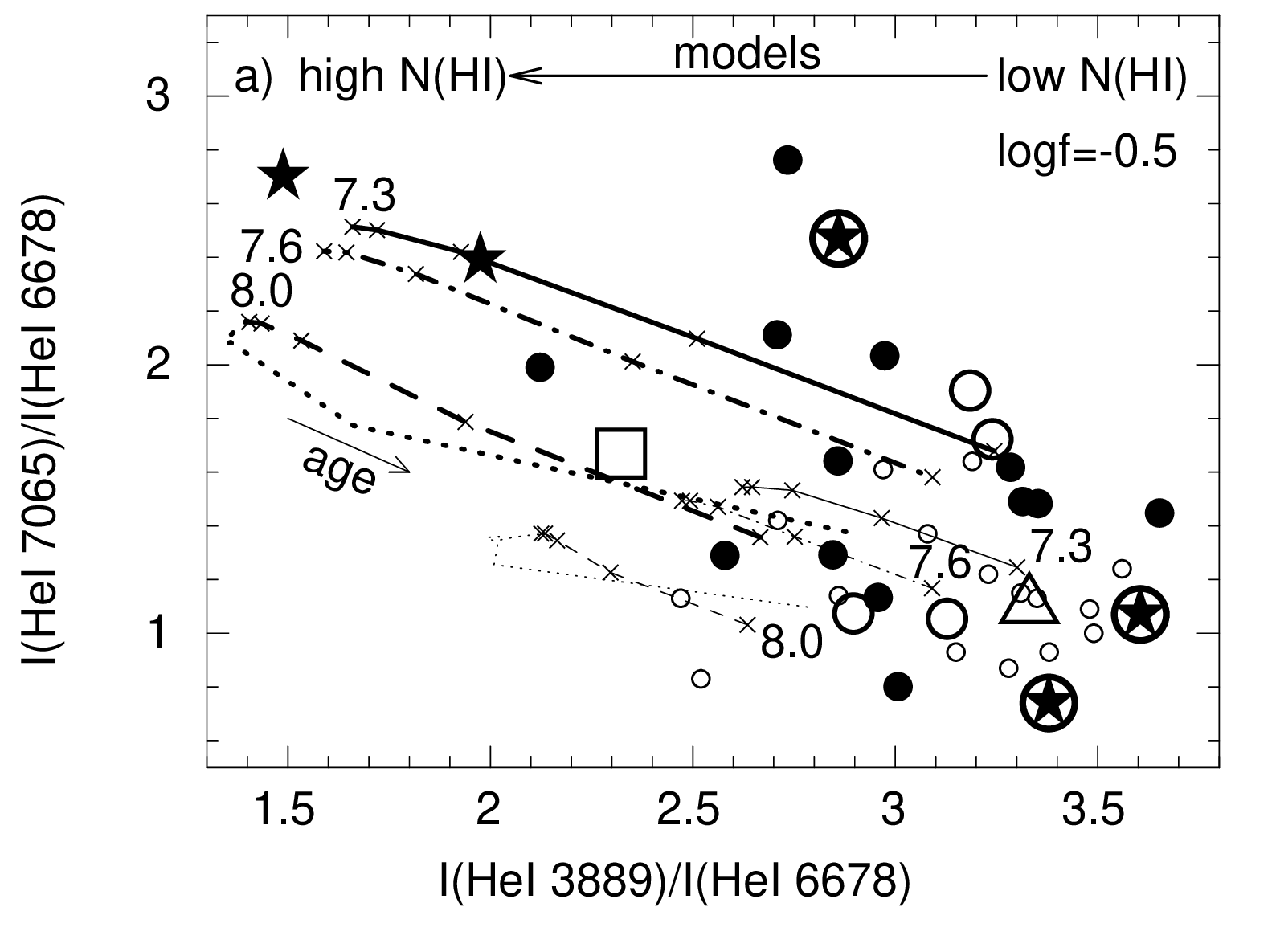}
\hspace{0.5cm}\includegraphics[angle=-90,width=0.40\linewidth]{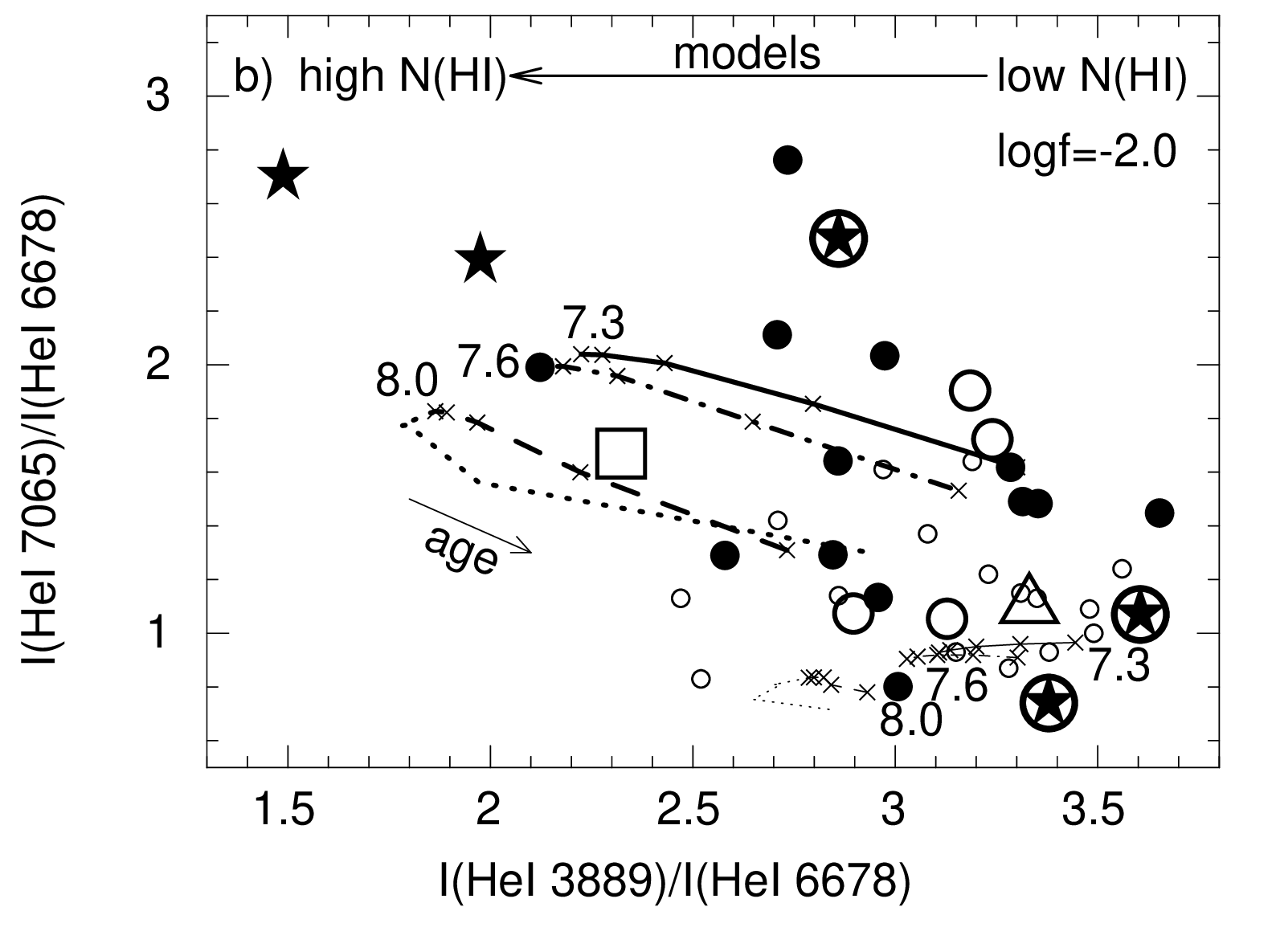}
\caption{The $I$(He {\sc i} $\lambda$7065)/$I$(He {\sc i} $\lambda$6678) -- 
$I$(He {\sc i} $\lambda$3889)/$I$(He {\sc i} $\lambda$6678) diagram
for (a) $\log f$ = $-$0.5 and (b) $\log f$ = $-$2.0. Symbols
and lines are the same as in Fig.~\ref{fig6}.
The LyC leaker J1333+6246 \citep{I16b} with noisy
He {\sc i} $\lambda$6678 and 
$\lambda$7065 lines in its SDSS spectrum is not shown.
}
\label{fig8}
\end{figure*}

\section{O$_{32}$ and helium line intensities as diagnostics for Lyman continuum leakage in CSFGs}\label{sec:diagnostics}

One of the important questions of contemporary cosmology is whether SFGs can
lose their ionizing radiation and, if so, in which quantities. The answer 
to that question has direct consequences  
on the problem of the reionization of the Universe. It has been suggested that
dwarf SFGs at redshifts 5 -- 10 can be the 
main contributors of the intergalactic 
medium ionization, if the escape fraction of their ionizing 
radiation is $\ga$ 10 percent \citep{O09,R13,Robertson15}. Such galaxies
could have large O$_{32}$ because of their low column densities of neutral gas.
However, O$_{32}$ depends also on metallicity, the starburst age, 
and the ionization parameter which is higher in denser, compact, and more 
luminous SFGs. Because of these other dependences,
\citet{S15} have concluded from various emission-line diagnostic diagrams that 
the escape fraction from photoionized H {\sc ii} regions
cannot exceed 10 percent, even with the highest observed O$_{32}$. If shocks 
are present, they can enhance the 
[O {\sc ii}] $\lambda$3727\AA\ emission line, reducing O$_{32}$ as compared to 
the value expected for a density-bounded photoionized H {\sc ii} region
\citep{JO13,S15}.

Therefore, other diagnostics in the optical range are needed which are based not
only on O$_{\rm 32}$. We propose
to use the He {\sc i} emission lines, that are sensitive to the electron 
number density and the optical depth, to find density-bounded H {\sc ii} 
regions. Helium 
emission, in contrast to the collisionally excited oxygen lines, is mainly 
produced via recombination. However, in the presence of a significant  
electron number density and a large optical depth in some He~{\sc i} 
emission lines, there can also be contributions from collisional excitation 
and fluorescent enhancement processes. By examining the intensity ratios of 
several He~{\sc i} emission lines, we are able to constrain both the 
electron number density and the optical depth in the H~{\sc ii} region.

In the optical range, the optimal set of He {\sc i} emission lines for 
constraining those two quantities consists of a set of three 
lines: the $\lambda$3889\AA, $\lambda$6678\AA\ and 
$\lambda$7065\AA\ lines. All three lines are sensitive to collisional
excitation, with the most sensitive one being the He~{\sc i} $\lambda$7065\AA\ line
\citep*{B99,B02,I13}. Additionally, the He~{\sc i} $\lambda$3889\AA\ 
and $\lambda$7065\AA\ emission lines are also 
sensitive to fluorescent enhancement.
However, the He~{\sc i} $\lambda$6678\AA\ emission line intensity is 
nearly insensitive to this effect.
While collisional excitation enhances all
lines, the effect of optical depth is different.
Thus, a non-negligible optical depth of the He~{\sc i} $\lambda$3889\AA\ 
transition decreases its intensity.
The problem, however, with this line is that it
is blended with the hydrogen H8 $\lambda$3889\AA\ emission line. Therefore, we 
have subtracted the hydrogen line from the blend, adopting its intensity 
to be equal to 0.107 
that of the H$\beta$ emission line for the electron temperature 
$T_{\rm e}$ = 15000 - 20000K \citep{SH95}. As for the He~{\sc i} 
$\lambda$7065\AA\ emission line, its intensity is strongly increased because of 
both collisional excitation and fluorescence. It is seen in Fig.~\ref{fig3} 
that the He~{\sc i} $\lambda$7065 line in J0159+0751, J1032+4919 and J1205+4551
is much stronger than the He~{\sc i} $\lambda$6678 line, indicating higher  
densities and/or optical depths in these galaxies. On the other
hand, in J1355+4651 and J1608+3528, these two lines are comparable in strength,
implying lower $N_{\rm e}$(He {\sc i}) and/or optical depths.

\citet*{I14b} have shown that the electron number density $N_{\rm e}$(He {\sc i})
is correlated with $N_{\rm e}$(S~{\sc ii}) in high-excitation H {\sc ii}
regions, although in relatively dense regions, $N_{\rm e}$(He {\sc i}) is 
somewhat larger than $N_{\rm e}$(S~{\sc ii}). Since $N_{\rm e}$(He {\sc i}) is
a characteristic of the entire H {\sc ii} region, while $N_{\rm e}$(S~{\sc ii})
is related only to its outer zone and the photodissociation region (PDR) around
it, we may conclude that the central part of the H {\sc ii} region is denser
than its outer part. In this paper, we use the 
He {\sc i} $\lambda$3889/$\lambda$6678 and $\lambda$7065/$\lambda$6678
flux ratios, which depend on both the $N_{\rm e}$(He {\sc i}) and 
$\tau$(He {\sc i} $\lambda$3889), to check whether this conclusion holds for our
five CSFGs as well. For this, we adopt the analytical 
approximations of the He~{\sc i} line
fluxes by \citet{I13}, based on the He~{\sc i} emissivities of \citet{P12,P13}.
These approximations depend on three parameters, the electron temperature, 
the electron number density $N_{\rm e}$(He {\sc i}) and the optical
depth $\tau$(He~{\sc i} $\lambda$3889). For the electron 
temperature, we adopt $T_{\rm e}$(O~{\sc iii}) (Table \ref{tab5}), while
$N_{\rm e}$(He {\sc i}) and $\tau$(He {\sc i} $\lambda$3889) are derived from the
extinction-corrected He {\sc i} $\lambda$3889/$\lambda$6678 and 
$\lambda$7065/$\lambda$6678 flux ratios (Table \ref{tab4}).

The results of our calculations are shown in Table~\ref{tab7}, where we 
present the 
derived $N_{\rm e}$(He {\sc i}) and $\tau$(He {\sc i} $\lambda$3889) and 
compare the observed and calculated He {\sc i} $\lambda$3889/$\lambda$6678 and 
$\lambda$7065/$\lambda$6678 flux ratios, 
together with the 
low-density case B ratios. By comparing $N_{\rm e}$(He {\sc i}) in 
Table~\ref{tab7} with $N_{\rm e}$(S~{\sc ii}) in
Table \ref{tab5}, we find that $N_{\rm e}$(He {\sc i}) is, in 
four out of five CSFGs, higher than $N_{\rm e}$(S~{\sc ii}), 
and in one object, J1205+4551, they are
comparable. This 
confirms the result of \citet{I14b} that,
on average, $N_{\rm e}$(He {\sc i}) is higher
than $N_{\rm e}$(S~{\sc ii}) in H {\sc ii} regions. Furthermore,
in the three CSFGs with a strong He {\sc i} $\lambda$7065 line,
J0159+0751, J1032+4919 and J1205+4551,  
the electron number density 
and optical depth are considerably higher than in the remaining two galaxies,
J1355+4651 and J1608+3528. Since, according to photoionization models of 
H {\sc ii} regions, $\tau$(He~{\sc i} $\lambda$3889) increases with
increasing column density of the neutral gas, we conclude that the former
three galaxies are more optically thick to the LyC radiation 
than the two later ones, although the optical depth in J0159+0751
is considerably smaller than those in J1032+4919 and J1205+4551 
(Table \ref{tab7}).

We further produce a grid of photoionized H {\sc ii} region models using the
{\sc Cloudy} code c13.04 \citep{F98,F13} with fixed neutral hydrogen column 
densities $N$(H~{\sc i}), ranging from 10$^{17}$ to 10$^{19}$ cm$^{-2}$, with the 
lowest and highest values corresponding to
two limiting cases of the optical depth: a low optical depth of $\sim$0.7 
in the Lyman continuum and a high escape fraction $f_{\rm esc}^{\rm LyC}$ $\ga$ 20 
percent, and a high optical 
depth of $\sim$70 in the Lyman continuum and a negligible LyC escape fraction.

It has empirically been found from {\sl HST} observations that the 
escape fraction $f_{\rm esc}^{\rm LyC}$ of ionizing radiation in low-$z$ leakers 
increases with increasing O$_{32}$ \citep{I16b}. We wish here to investigate whether this
empirical relation can also be predicted by photoionized H~{\sc ii} region models, i.e. are  
high modelled O$_{32}$ ratios reliable indicators of the escaping LyC radiation? 
Similar studies have been conducted by \citet{JO13}, \citet{NO14} 
and \citet{S15}, with somewhat contradictory and inconclusive results. 
\citet{S15} used various emission-line diagnostics of a large sample of
BCDs from the SDSS and found that their properties can be fully explained by
ionization-bounded H~{\sc ii} region models with a low LyC escape fraction,
including galaxies with the highest O$_{32}$, if the ionizing radiation flux 
is purely of stellar origin. On the other hand, \citet{JO13} and \citet{NO14} 
suggested that high
O$_{32}$ may be a good indicator of escaping LyC radiation. However, 
although some models of \citet{NO14} do predict higher O$_{32}$ at higher 
$f_{\rm esc}^{\rm LyC}$, they also have models that allow escaping LyC radiation 
with any O$_{32}$. Furthermore, modelled O$_{32}$ ratios in high-$z$
Ly$\alpha$ emitting galaxies are a factor $\sim$ 2--5 too high for their 
ionization parameters \citep[see fig. 11 in ][]{NO14}. 
Thus, by itself, the modelled O$_{32}$ is a poor indicator of $f_{\rm esc}$(LyC). 
It should be combined with some observational data, such as
metallicity, various emission-line flux ratios, and the H$\beta$ equivalent
width EW(H$\beta$) which is related to the starburst age and hence to  
the hardness of the 
ionizing radiation. Used together, these data would 
better constrain the physical conditions in the H~{\sc ii} region. 

The modelled O$_{32}$ ratio at a fixed ionization parameter can be 
decreased to better fit observed ratios if other sources of ionizing radiation, 
e.g. radiative shocks, are considered \citep{JO13,S15}. Unfortunately, in its 
present state, the {\sc Cloudy} code does not include self-consistently both 
the stellar radiation
and the radiation from radiative shocks. Furthermore, the fast radiative shock 
models in existence consider only shocks propagating through the neutral gas 
\citep{A08}, while in CSFGs, shocks
are also likely to propagate through the ionized medium. Additionally,
the modelled O$_{32}$ ratio for a fixed ionization parameter decreases  
when the starburst age increases \citep{JO13}.

Keeping in mind all these caveats, we have generated a grid of {\sc Cloudy} 
models to check the reliability of the modelled O$_{32}$ ratio as an indicator 
of LyC leakage. There are many input parameters in the {\sc Cloudy} models. 
Some of the
parameters can be obtained directly from observations. The chemical composition
can be derived from the emission-line intensities in the galaxy spectrum.
The rate of ionizing photon production $Q$ and the shape of the ionizing
radiation spectrum are determined from the 
H$\beta$ luminosity and the starburst age. The latter quantity can be derived, 
for an instantaneous burst, from the H$\beta$ equivalent width EW(H$\beta$). The
electron number density $N_{\rm e}$ is derived from the observed
intensities of [S {\sc ii}]$\lambda$6717, 6731 or from the He {\sc i} emission 
lines. The determination of other input parameters, for instance, 
the inner radius of the H~{\sc ii} region, the turbulent velocity, and the 
filling factor $f$, is more complicated. They can be derived by comparing 
the observed and modelled emission-line intensities for each galaxy.

Our aim here is to study general statistical trends 
of the galaxy sample. Therefore, we have calculated a grid of models for
three different values of the oxygen abundance,
12+logO/H = 7.3, 7.6 and 8.0, and a fixed ionization parameter. For elements 
heavier than helium, we have adopted their
abundances relative to that of oxygen to be the abundance ratios
given by \citet{IT99} and \citet{I06}. The helium abundance is chosen from the 
relation 
between the helium mass fraction and oxygen abundance of \citet{I14b}. We  also 
included dust with Orion-like properties, provided by the {\sc Cloudy} code, 
and scaling it according to the oxygen 
abundance. We adopt $Q$ = 10$^{53}$ s$^{-1}$, which corresponds to the average
observed H$\beta$ luminosity in our galaxies, and a shape of the ionizing 
radiation spectrum corresponding to the Starburst99 model with an age of 2.0
Myr. 
For the inner radius of the H {\sc ii} region, we have adopted the value
$R_{\rm in}$ = 50 pc. Reducing $R_{\rm in}$ by a factor of 10 would result in
increasing O$_{32}$ by a factor of up to $\sim$ 1.5 in models with 
$N$(H~{\sc i}) $\la$ 10$^{18}$ cm$^{-2}$, but the He {\sc i} flux ratios would be
changed by not more than 10 percent in models with any $N$(H~{\sc i}).
The models were calculated for two values of the filling factor $f$ with
log~$f$ = $-0.5$ and $-2.0$  and two electron number densities 
$N_{\rm e}$ = 100 and 1000 cm$^{-3}$. We also set the turbulent velocity
to 0.

To investigate the effect of the hardness of the ionizing radiation 
at a given metallicity, we have calculated additionally, for an oxygen 
abundance 12+logO/H = 8.0, a set of models with starburst ages in the range 
1 -- 6 Myr, characterized by an equivalent width EW(H$\beta$) in the range 
$\sim$~50 -- 500\AA\ (thick dotted lines for $N_{\rm e}$ = 1000 cm$^{-3}$ and 
thin dotted lines for $N_{\rm e}$ = 100 cm$^{-3}$ in Fig.~\ref{fig8}).
The other parameters remained the same.

We wish to find the models which fit simultaneously the 
[O~{\sc iii}]$\lambda$5007/H$\beta$ emission-line ratio, the O$_{32}$ ratio, 
and the He~{\sc i} line ratios.
The results of our model calculations and their comparison with the data
are shown in Figs.~\ref{fig6}--\ref{fig8}. 
The diagrams [O~{\sc iii}]$\lambda$5007/H$\beta$ -- O$_{32}$ and
$\lambda$7065/$\lambda$6678 -- 
O$_{32}$ for models with log $f$ =$-$0.5 and log $f$ =$-$2.0,
corresponding to the logarithm of the ionization parameter in the range 
$\sim$ $-$1.5 - $-$2.5, are presented in 
Figs.~\ref{fig6}a -- \ref{fig6}b and Figs.~\ref{fig7}a -- \ref{fig7}b,
respectively. We note that, for a given metallicity and a given electron number
density, the [O~{\sc iii}]$\lambda$5007/H$\beta$ ratio changes only slowly 
with $N$(H~{\sc i}), while O$_{32}$ undergoes large variations. 
It is seen that the location of all extreme CSFGs (filled stars)
can be explained by models with log $f$ =$-$0.5 and high $N$(H~{\sc i}) 
(Fig.~\ref{fig6}a and \ref{fig7}a). 
Alternatively, models with much lower filling factors 
(log $f$ =$-$2.0), and correspondingly, with lower ionization parameters, 
are needed to fit observations with lower modelled $N$(H~{\sc i}) 
(Fig.~\ref{fig6}b and \ref{fig7}b). However, in the latter case, the models
are unable to reproduce the observed [O~{\sc iii}]$\lambda$5007/H$\beta$ ratios 
(Fig.~\ref{fig6}b), predicting too low values for the galaxy metallicities 
(Table \ref{tab5}). Furthermore, the models fail to reproduce the observed line
intensity ratios for three out of the five extreme CSFGs in Fig.~\ref{fig7}b.
Therefore, we conclude that the models with low log $f$ =$-$2.0, and hence 
with low ionization parameters, are not 
likely good fits to our extreme CSFGs.

For comparison, we have also plotted other data sets in those Figures. The 
first set consists of the LyC leakers studied by \citet{I16a,I16b} (large open 
circles). We find that, in Figs. \ref{fig6} -- \ref{fig7}, these objects 
lie in the region corresponding to high $N$(H~{\sc i}), irrespective 
of the value of log $f$, in apparent disagreement with the relatively
high LyC escape fraction observed for these objects.
Similarly to the extreme CSFGs, 
the [O~{\sc iii}]$\lambda$5007/H$\beta$ ratios
of $\sim$ 6 in the LyC leakers with oxygen abundances 7.8 -- 8.0 are well 
reproduced by models with log $f$ =$-$0.5 while models with $f$ =$-$2.0
predict too low values.
The second data set shown (filled circles) consists of a sample of 
high-excitation CSFGs with O$_{32}$ $\sim$ 5 -- 20 and oxygen abundances
in the range $\sim$7.3 -- 7.8, observed with the LBT 
(Izotov et al. 2017, in preparation). The range of the 
$\lambda$7065/$\lambda$6678 ratios of the galaxies in the second set is 
similar to that of 
CSFGs with extremely high O$_{32}$. The galaxies are also located in the region
corresponding to models with high $N$(H~{\sc i}). The third set includes
Ly$\alpha$ emitting GPs observed with the {\sl HST}/COS 
\citep[small open circles, ][]{Y17}.
These objects, with oxygen abundances of $\sim$ 8.0, 
are characterized by O$_{32}$ ratios in the lowest range, 
$\sim$ 2 -- 10, and they are the most deviant from model predictions.
They also have lower [O~{\sc iii}]$\lambda$5007/H$\beta$ ratios and
lower H$\beta$ equivalent widths EW(H$\beta$), indicating higher burst ages
compared to extreme CSFGs and LyC leakers. 

To investigate the effects of age and hence varying the hardness
of ionizing radiation we show by dotted lines in 
Fig.~\ref{fig6} the dependences for optically thick models with a fixed oxygen 
abundance of 8.0, and $N_{\rm e}$ of 1000 cm$^{-3}$ (thick dotted line) and of 
100 cm$^{-3}$ (thin dotted line), and starburst ages varying from 1 Myr to 6 
Myr in the direction shown by the arrow. These models indicate that 
O$_{32}$ strongly depends not only on the ionization parameter, but also on the
hardness of the ionizing radiation. The range of predicted 
[O~{\sc iii}]$\lambda$5007/H$\beta$ ratios is sufficiently large, 
extending down to $<$ 3 for starburst ages $\geq$ 5 Myr, to account for the 
observed values. However, the modelled values of O$_{32}$ remain too high for 
a fixed [O~{\sc iii}]$\lambda$5007/H$\beta$ ratio, and the models indicate high
$N$(H~{\sc i}) for the Ly$\alpha$ emitting GPs. 

In Fig. \ref{fig7}, the modelled O$_{32}$ span a large range of values (dotted 
lines) and can reproduce the observed O$_{32}$ in the Ly$\alpha$ emitting GPs 
\citep[small open circles, ][]{Y17}. However, the lowest O$_{32}$ ($<$~1) 
correspond to large starburst ages ($\geq$~5 Myr) and thus to low 
EW(H$\beta$) ($<$ 100\AA). This is inconsistent with the high 
EW(H$\beta$) ($>$ 100\AA) and O$_{32}$(~$\geq$~2), observed by \citet{Y17} in 
most of the Ly$\alpha$ emitting GPs.

Finally, we show respectively by an open square and an open triangle in 
Figs.~\ref{fig6} -- \ref{fig7} the BCDs SBS~0335$-$052 with the oxygen 
abundance of $\sim$ 7.3 \citep{I09} and Tol~1214$-$277 with the oxygen 
abundance of $\sim$ 7.5 
\citep*{I01}. Their [O~{\sc iii}]$\lambda$5007/H$\beta$ ratios are well
reproduced by the models with log $f$ =$-$0.5 (Fig. \ref{fig6}a).
These two objects have been observed by \citet{TI97} in the UV range 
with the {\sl HST}. The Ly$\alpha$ line in the UV spectrum of SBS 0335$-$052 is 
in absorption. The neutral hydrogen column density derived from the
line profile is very high, 7$\times$10$^{21}$ cm$^{-2}$ \citep{TI97},
implying a very low LyC escape fraction. On the other hand, strong Ly$\alpha$  
emission is detected in the UV spectrum of Tol 1214$-$277, suggesting a high 
LyC escape fraction. Yet, in spite of that difference, both galaxies with
the electron number densities $N_{\rm e}$ $\sim$ 200 cm$^{-3}$ in their
H {\sc ii} regions are characterised by similar O$_{32}$ $\sim$ 15
\citep{I01,I09}. 

The above discussion implies that O$_{32}$ derived from photoionized
H~{\sc ii} region models are in general too high compared to the observed ones 
for a fixed ionization parameter and metallicity.
Thus it cannot be a sure indicator of 
escaping Ly$\alpha$ radiation, nor can it indicate escaping LyC radiation and 
density-bounded H {\sc ii} regions with certainty. The possible reason for the
disagreement between modelled and observed O$_{32}$ could be due to the fact 
that the models include only ionizing radiation from stars. If additional 
sources of gas ionization and
heating are included in the model, e.g. radiative shocks producing extreme UV 
radiation or sources producing X-ray emission, then O$_{32}$ would be changed.
Furthermore, we have considered only homogeneous H~{\sc ii} region models with a
constant $N_{\rm e}$, although density inhomogeneities may play a role.

More definite conclusions about LyC leakage can however be drawn from 
the He~{\sc i} emission-line flux ratios. Fig. \ref{fig8} shows the 
$\lambda$7065/$\lambda$6678 -- $\lambda$3889/$\lambda$6678 diagram, where we 
display the modelled relations between these line ratios
for two values of the electron number density and for three values
of the oxygen abundance. 
There is a clear dependence of the relations
on metallicity caused presumably by varying electron temperatures
in H~{\sc ii} regions, the temperatures being higher for lower metallicities.
The He~{\sc i} $\lambda$7065/$\lambda$6678
ratio also increases with increasing $N_{\rm e}$ because the $\lambda$7065 
transition is much more sensitive to collisional enhancement compared to the 
$\lambda$6678 transition. On the other hand, the $\lambda$3889 transition
is less sensitive to collisional enhancement compared to the 
$\lambda$6678 transition, therefore the He~{\sc i} $\lambda$3889/$\lambda$6678
ratio decreases with increasing $N_{\rm e}$.

The locations of the two out of three extreme CSFGs (non-encircled stars) with 
the highest  
$\lambda$7065/$\lambda$6678 ratios correspond to high column densities 
($N$(H~{\sc i})$\ga$10$^{19}$ cm$^{-2}$). On the other hand, the 
three remaining CSFGs (encircled stars)
are located in the region corresponding to low column densities 
($N$(H~{\sc i})$\la$10$^{17.5}$ cm$^{-2}$). This implies that the LyC escape
fractions in these three galaxies may be high and that follow-up 
Ly$\alpha$ observations may be warranted. 

It is interesting that the locations of the four LyC leakers 
at $z$ $\sim$ 0.3 in Fig. \ref{fig8}a (large open circles) and of 
the Ly$\alpha$ emitting GPs (small open circles)
are consistent with the low $N$(H~{\sc i}) predicted by the models with
high log $f$. The $N$(H~{\sc i}) for the Ly$\alpha$ emitting GPs are predicted 
to be lower than 10$^{19}$ cm$^{-2}$.
However, we note that  the He~{\sc i} $\lambda$6678\AA\ and $\lambda$7065\AA\ 
line fluxes in LyC leakers measured from the SDSS spectra are somewhat
uncertain because of their faintness and their location in a wavelength range
that contains numerous residuals of night sky emission lines.
Our results are somewhat
contradictory with the higher neutral hydrogen column densities
derived by \citet{Y17}, using radiative transfer models for the Ly$\alpha$ line
in the Ly$\alpha$ emitting GPs. Those authors 
derived column densities $N$(H~{\sc i}) in the range 10$^{16}$ -- 10$^{20.6}$
cm$^{-2}$, with an average value of more than 10$^{19}$ cm$^{-2}$. Their
sample includes the five LyC leakers of  \citet{I16a,I16b}. For one LyC leaker,
J1503$+$3644, \citet{Y17} derived a low $N$(H~{\sc i}) = 10$^{16.8}$ cm$^{-2}$,
but the best fit of $N$(H~{\sc i}) for the remaining LyC leakers 
is $\ga$ 10$^{19}$ cm$^{-2}$, larger than the value that permits LyC escape. 
Therefore, \citet{Y17} suggested that LyC emission in these galaxies
probably escapes through holes in the interstellar medium with much lower 
$N$(H~{\sc i}).

As for the objects in the sample of CSFGs
with O$_{32}$ $\sim$ 5 -- 20 observed with the LBT (filled circles), they are  
distributed in a region spanning a broad range of $\lambda$3889/$\lambda$6678 
and $\lambda$7065/$\lambda$6678 line ratios, with a significant fraction 
of the galaxies characterized by low $N$(H~{\sc i}) (Fig. \ref{fig8}a), 
despite the fact that H~{\sc ii} regions with high $\lambda$7065/$\lambda$6678 
line ratios in some of these galaxies are very dense.
It is also interesting to note that the BCD SBS 0335$-$052 with 
its Ly$\alpha$ line in absorption is characterized, as expected, by a    
high $N$(H~{\sc i}), while the BCD Tol 1214$-$277 with its Ly$\alpha$ line in 
emission is characterized by a low $N$(H~{\sc i}).

Somewhat different conclusions are drawn if we 
compare the observations
with models characterized by a low log $f$ = $-$2.0. High 
$N$(H~{\sc i}) are predicted for many galaxies, including two LyC leakers, 
if a low electron number density $N_{\rm e}$ is adopted. In this
case, the observed [O~{\sc iii}]$\lambda$5007/H$\beta$ ratios are 
higher than those predicted by the models (Fig. \ref{fig6}b). Furthermore, 
the observed electron number densities $N_{\rm e}$(S~{\sc ii}) of most LyC leakers 
\citep{I16a,I16b} and of many Ly$\alpha$ emitting GPs are several times 
greater than 100 cm$^{-3}$, although these values are somewhat uncertain because of large errors. We therefore conclude that the models with
a low log $f$ = $-$2.0 and a low $N_{\rm e}$ are less likely to fit the data 
than those with high log $f$ = $-$0.5 and high $N_{\rm e}$. More accurate 
electron number densities are needed to draw more definite conclusions on the 
escaping ionizing radiation in the above galaxies.

To investigate the effect of varying hardness of radiation, 
we show by dotted lines in Fig. \ref{fig8} the same optically thick
models with various starburst ages, as in Figs. \ref{fig6} -- \ref{fig7}.
They reveal an interesting behaviour. For starburst ages $<$ 4 Myr, the He~{\sc i}
emission-line ratios depend weakly on starburst age. However, at larger ages,
the He~{\sc i} $\lambda$3889/$\lambda$6678 ratio increases with starburst
age, while the He~{\sc i} $\lambda$7065/$\lambda$6678 ratio decreases.
This behaviour can be understood as a softening of the ionizing radiation, 
resulting in a decrease of the size of the He$^+$ zone and thus in a decrease
of the optical depth in the He~{\sc i} $\lambda$3889 transition. However, all
galaxies considered here have a high EW(H$\beta$) and thus a younger burst age,
consistent with the case where the He~{\sc i}
emission-line ratios vary weakly with starburst age.

Thus, based admittedly on a small sample of galaxies 
with extreme O$_{32}$, our investigation has shown that the proposed
He {\sc i} emission flux ratios in the optical range can be useful
diagnostics for finding Ly$\alpha$ and LyC leaking galaxies.

\section{Conclusions}\label{sec:conclusions}

In this paper we present Large Binocular Telescope (LBT) 
spectrophotometric observations of five compact star-forming galaxies (CSFGs)
with extremely high O$_{32}$ = 23 -- 43. CSFGs with such high O$_{32}$ are
 thought to be good candidates for leaking LyC radiation. As such,    
their high-redshift counterparts are thought to be the main agents of 
the reionization of the early Universe. Our goal here is to study the  
physical conditions and chemical composition of these extreme CSFGs and to 
investigate whether they can indeed be candidates for LyC leaking galaxies. 
Our main results are as follows.

1. All spectra show strong emission lines implying the presence of a very 
young stellar population. 
This is supported by very high equivalent widths EW(H$\beta$) 
of the H$\beta$ emission line with values 370 -- 520\AA, corresponding to a starburst
age $<$ 3 Myr. The electron number densities $N_{\rm e}$, derived from the
[S~{\sc ii}]$\lambda$6717/$\lambda$6731 flux ratio, 
are particularly high. They are 190 -- 260 cm$^{-3}$
in two CSFGs and higher, 470 -- 640 cm$^{-3}$, in the three others.

2. A strong [O {\sc iii}] $\lambda$4363\AA\ emission line is detected in all
objects, allowing element abundance determination
 by the direct $T_{\rm e}$ method. We
find high electron temperatures $T_{\rm e}$(O {\sc iii}) $\sim$ 17000 -- 21000K 
and low oxygen abundances 12 + logO/H in the range 7.46 -- 7.79. The Ne/O,
S/O, Cl/O, Ar/O and Fe/O abundance ratios in all five galaxies are similar
to those found in low-metallicity blue compact dwarf (BCD) galaxies.
The N/O abundance ratios in three CSFGs are low and also consistent with
values found for low-metallicity BCDs. On the other hand, in the two remaining 
galaxies, the nitrogen abundance is higher by up to 0.8 dex, which is too
high to be explained by the 
chemical enrichment of homogeneous H~{\sc ii} regions,
with a constant electron number density. 
Nitrogen-enriched clumps in inhomogeneous H~{\sc ii} regions need to 
be invoked to explain such high N/O abundance ratios.  

3. A broad H$\alpha$ emission line is detected in the 
spectra of all CSFGs, indicating
the presence of fast dynamical processes, possibly related to young 
supernova remnants in the early stages of their evolution. 
The full widths at half maximum of the broad component correspond to 
velocities of $\sim$ 1800 -- 2000 km s$^{-1}$. Such rapid ionized 
gas motions may facilitate the escape of the resonant Ly$\alpha$ emission 
from the galaxy. In one of the CSFGs,
J1205+4551, the high-ionization emission line [Ne {\sc v}] $\lambda$3426\AA\
is detected, implying the presence of shocks possibly produced by SNe.

4. Various He {\sc i} emission lines are present in the 
spectra of the observed CSFGs.
We find that the He {\sc i} $\lambda$7065/$\lambda$6678 flux ratios in three
out of five CSFGs are unusually high, indicating high electron number densities
$N_{\rm e}$(He {\sc i}) and/or high optical depths $\tau$($\lambda$3889) in
the He {\sc i} $\lambda$3889 emission line. This is consistent with the high
electron number densities derived from the [S {\sc ii}] emission lines.
The He {\sc i} $\lambda$7065/$\lambda$6678 flux ratios in the remaining
two CSFGs are considerably lower, corresponding to a 
lower $N_{\rm e}$(He {\sc i}) and/or
$\tau$($\lambda$3889). 

5. We investigate whether the LyC can leak out of these CSFGs by constraining 
their neutral gas column density $N$(H~{\sc i}) through {\sc Cloudy} models, 
using O$_{32}$ and the He~{\sc i} emission-line flux ratios.
We find that the modelled O$_{32}$ is not a certain indicator for LyC leakage 
because it depends also on other parameters, such as the ionization parameter
and the hardness of ionizing radiation, the latter depending in turn on
the starburst age. 
Therefore, it alone cannot be used with H~{\sc ii} region models for 
estimation of the neutral gas column density and the LyC escape fraction. The 
only reliable way for the determination of the $f_{\rm esc}^{\rm LyC}$ is to obtain 
a well-defined empirical relation, based on a large sample, between 
$f_{\rm esc}^{\rm LyC}$ and O$_{32}$, both derived directly from observations.
On the other hand, the modelled He {\sc i} $\lambda$3889/$\lambda$6678 and
$\lambda$7065/$\lambda$6678 emission-line flux ratios are better indicators of 
possible LyC leakage. We find that in two CSFGs with high 
$\lambda$7065/$\lambda$6678 flux ratios, the neutral gas column density 
$N$(H~{\sc i}) is high, $\ga$10$^{19}$ cm$^{-2}$, implying a low LyC escape 
fraction. However, in the three remaining galaxies $N$(H~{\sc i})
can be below 10$^{17.5}$ cm$^{-2}$, suggesting that these galaxies can
lose a large fraction of their ionizing radiation, up to $\ga$ 20 percent.

\section*{Acknowledgements}

Funding for the SDSS-III has been provided by the Alfred P. Sloan Foundation, the Participating Institutions, the National Science Foundation, the U.S. Department of Energy, the National Aeronautics and Space Administration, the Japanese Monbukagakusho, the Max Planck Society, and the Higher Education Funding Council for England. The SDSS Web Site is http://www.sdss.org/.
GALEX is a NASA mission managed by the Jet Propulsion Laboratory.
This research has made use of the NASA/IPAC Extragalactic
Database (NED) which is operated by the Jet
Propulsion Laboratory, California Institute of Technology,
under contract with the National Aeronautics and Space
Administration. This publication makes use of data products
from the Wide-field Infrared Survey Explorer, which is
a joint project of the University of California, Los Angeles,
and the Jet Propulsion Laboratory/California Institute of
Technology, funded by the National Aeronautics and Space
Administration.

\input{ref.tex}
\bsp

\label{lastpage}

\end{document}

%% file: tab1_1.tex
\begin{table*}
\caption{Observed characteristics of galaxies \label{tab1}}
\begin{tabular}{lcccccccccccccc} \hline
Name      &R.A.(J2000)&Dec.(J2000) &O$_{32}$$^{\rm a}$&  $z$ & $g$&$FUV$&$NUV$&$W1$&$W2$&$W3$&$W4$  \\
          &           &            &       &      &(mag)&(mag)&(mag)&(mag)&(mag)&(mag)&(mag) \\ \hline
J0159+0751&01:59:52.75&+07:51:48.90&39&0.0611&19.47&20.64&20.43&17.39&15.88&11.54&8.52 \\
J1032+4919&10:32:56.72&+49:19:47.24&24&0.0442&18.67&20.26&20.25&16.60&15.08&10.50&7.42 \\
J1205+4551&12:05:03.55&+45:51:50.94&23&0.0654&19.79&21.04&20.58&15.14&13.55&~~9.91&7.80 \\ 
J1355+4651&13:55:25.64&+46:51:51.34&23&0.0281&19.22&20.54&20.66&17.86&17.39& ... & ... \\ 
J1608+3528&16:08:10.36&+35:28:09.30&43&0.0327&18.59&20.01&20.42&18.02&17.06&11.56&8.77 \\ 
\hline
  \end{tabular}

\hbox{$^{\rm a}$derived from the LBT spectra.}

  \end{table*}

%% file: tab2.tex
\begin{table*}
\caption{Global characteristics of galaxies \label{tab2}}
\begin{tabular}{lcccccc} \hline
Name      &log $M_\star$/M$_\odot$&log $L$(H$\beta$)&SFR&$M_g$&$M_{FUV}$&12+logO/H  \\
          &                 &(log erg s$^{-1}$)&(M$_\odot$ yr$^{-1}$)&(mag) &(mag) &    \\ \hline
J0159+0751&6.69&40.93&1.9&$-$18.73&$-$17.56&7.54 \\
J1032+4919&6.65&40.74&1.2&$-$18.01&$-$16.42&7.65 \\
J1205+4551&6.84&41.11&2.9&$-$18.56&$-$17.31&7.46 \\ 
J1355+4651&6.05&39.99&0.2&$-$16.68&$-$15.36&7.57 \\ 
J1608+3528&7.04&40.41&0.6&$-$18.43&$-$17.01&7.79 \\ 
\hline
  \end{tabular}
  \end{table*}

%% file: tab3.tex
\begin{table*}
\caption{Large Binocular Telescope observations \label{tab3}}
\begin{tabular}{lccccc} \hline
Name      &Date&Instrument&Exposure &Slit    &Airmass  \\
          &    &          &    (s)  &(arcsec)&         \\ \hline
J0159+0751&2016-01-02&MODS1&3200&60$\times$1.2&1.10 \\
J1032+4919&2016-02-12&MODS1&2400&60$\times$1.2&1.04 \\
J1205+4551&2016-04-10&MODS1&3200&60$\times$1.2&1.24 \\
J1355+4651&2013-06-08&MODS1&2400&60$\times$1.2&1.44 \\
J1608+3528&2013-06-08&MODS1&2400&60$\times$1.2&1.42 \\
\hline
  \end{tabular}
  \end{table*}

%% file: tab4_1.tex
\begin{table*}
\caption{Extinction-corrected emission-line fluxes \label{tab4}}
\begin{tabular}{lrrrrr} \hline
Line&\multicolumn{5}{c}{100$\times$$I$($\lambda$)/$I$(H$\beta$)} \\ \cline{2-6}
    &J0159+0751&J1032+4919&J1205+4551&J1355+4651&J1608+3528 \\ \hline
3426.85 [Ne {\sc v}]            &...~~~~~       &...~~~~~       &  0.72$\pm$0.11&...~~~~~       &...~~~~~        \\
3727.00 [O {\sc ii}]            & 16.15$\pm$0.55& 27.86$\pm$0.89& 18.26$\pm$0.59& 24.08$\pm$0.59& 16.85$\pm$0.62 \\
3750.15 H12                     &  4.43$\pm$0.26&  3.62$\pm$0.17&  4.42$\pm$0.19&  3.82$\pm$1.20&  4.12$\pm$0.37 \\
3770.63 H11                     &  5.10$\pm$0.26&  4.70$\pm$0.19&  4.78$\pm$0.20&  4.98$\pm$1.95&  4.96$\pm$0.36 \\
3797.90 H10                     &  6.83$\pm$0.29&  6.01$\pm$0.22&  6.19$\pm$0.23&  8.23$\pm$1.52&  6.49$\pm$0.40 \\
3819.64 He {\sc i}              &  1.29$\pm$0.12&  0.99$\pm$0.06&  1.19$\pm$0.07&...~~~~~       &  1.51$\pm$0.21 \\
3835.39 H9                      &  8.64$\pm$0.33&  7.86$\pm$0.27&  8.15$\pm$0.28&  8.57$\pm$1.31&  7.79$\pm$0.40 \\
3868.76 [Ne {\sc iii}]          & 56.68$\pm$1.78& 53.67$\pm$1.67& 24.25$\pm$0.76& 49.73$\pm$1.69& 60.19$\pm$1.90 \\
3889.00 He {\sc i}+H8           & 18.61$\pm$0.63& 16.41$\pm$0.52& 15.13$\pm$0.49& 20.97$\pm$1.04& 19.52$\pm$0.67 \\
3968.00 [Ne {\sc iii}]+H7       & 35.47$\pm$1.11& 33.85$\pm$1.04& 24.56$\pm$0.77& 33.75$\pm$1.38& 33.06$\pm$1.06 \\
4026.19 He {\sc i}              &  2.13$\pm$0.11&  1.97$\pm$0.08&  1.97$\pm$0.08&  2.56$\pm$0.26&  1.71$\pm$0.10 \\
4068.60 [S {\sc ii}]            &...~~~~~       &  0.79$\pm$0.06&  0.65$\pm$0.06&...~~~~~       &  0.61$\pm$0.02 \\
4076.35 [S {\sc ii}]            &...~~~~~       &  0.18$\pm$0.05&  0.30$\pm$0.06&...~~~~~       &...~~~~~        \\
4101.74 H$\delta$               & 27.71$\pm$0.86& 26.81$\pm$0.81& 27.50$\pm$0.84& 25.75$\pm$1.14& 23.65$\pm$0.76 \\
4120.84 He {\sc i}              &...~~~~~       &  0.27$\pm$0.04&  0.44$\pm$0.05&...~~~~~       &  0.66$\pm$0.09 \\
4143.76 He {\sc i}              &  0.72$\pm$0.07&  0.61$\pm$0.06&  0.57$\pm$0.05&...~~~~~       &  0.31$\pm$0.06 \\
4227.20 [Fe {\sc v}]            &  1.35$\pm$0.08&  0.78$\pm$0.05&  0.73$\pm$0.05&  1.42$\pm$0.27&  0.40$\pm$0.06 \\
4340.47 H$\gamma$               & 48.29$\pm$1.43& 47.56$\pm$1.39& 46.99$\pm$1.38& 44.63$\pm$1.46& 43.36$\pm$1.28 \\
4363.21 [O {\sc iii}]           & 22.90$\pm$0.68& 20.85$\pm$0.61& 12.87$\pm$0.38& 18.13$\pm$0.57& 18.96$\pm$0.56 \\
4387.93 He {\sc i}              &  0.89$\pm$0.09&  0.49$\pm$0.04&  0.49$\pm$0.05&  1.08$\pm$0.16&  0.60$\pm$0.05 \\
4471.48 He {\sc i}              &  4.12$\pm$0.15&  4.10$\pm$0.13&  4.02$\pm$0.13&  3.56$\pm$0.16&  3.59$\pm$0.12 \\
4658.10 [Fe {\sc iii}]          &...~~~~~       &  0.38$\pm$0.05&  0.58$\pm$0.05&...~~~~~       &  0.28$\pm$0.03 \\
4685.94 He {\sc ii}             &  1.58$\pm$0.09&  1.16$\pm$0.06&  3.24$\pm$0.11&  1.61$\pm$0.10&  1.89$\pm$0.08 \\
4712.00 [Ar {\sc iv}]+He {\sc i}&  4.54$\pm$0.15&  3.15$\pm$0.10&  2.15$\pm$0.05&  3.67$\pm$0.16&  4.95$\pm$0.16 \\
4740.20 [Ar {\sc iv}]           &  3.62$\pm$0.14&  2.37$\pm$0.08&  1.20$\pm$0.06&  3.16$\pm$0.14&  3.45$\pm$0.12 \\
4861.33 H$\beta$                &100.00$\pm$2.87&100.00$\pm$2.85&100.00$\pm$2.86&100.00$\pm$2.92&100.00$\pm$2.87 \\
4921.93 He {\sc i}              &  1.06$\pm$0.08&  1.09$\pm$0.05&  1.03$\pm$0.05&  0.92$\pm$0.09&  1.33$\pm$0.06 \\
4958.92 [O {\sc iii}]           &211.54$\pm$6.06&221.47$\pm$6.31&139.15$\pm$3.98&194.24$\pm$5.58&252.84$\pm$7.21 \\
4988.00 [Fe {\sc iii}]          &...~~~~~       &  0.57$\pm$0.05&...~~~~~       &...~~~~~       &...~~~~~        \\
5006.80 [O {\sc iii}]           &631.72$\pm$18.1&665.43$\pm$19.0&414.63$\pm$11.9&559.95$\pm$16.1&730.81$\pm$20.9 \\
5015.68 He {\sc i}              &  1.96$\pm$0.10&  1.78$\pm$0.07&  1.98$\pm$0.07&  1.48$\pm$0.07&  1.91$\pm$0.06 \\
5411.52 He {\sc ii}             &...~~~~~       &...~~~~~       &  0.32$\pm$0.03&...~~~~~       &...~~~~~        \\
5517.71 [Cl {\sc iii}]          &...~~~~~       &  0.17$\pm$0.03&...~~~~~       &...~~~~~       &...~~~~~        \\
5537.88 [Cl {\sc iii}]          &...~~~~~       &  0.13$\pm$0.03&...~~~~~       &...~~~~~       &...~~~~~        \\
5754.64 [N {\sc ii}]            &...~~~~~       &...~~~~~       &  0.32$\pm$0.03&...~~~~~       &...~~~~~        \\
5875.60 He {\sc i}              & 11.22$\pm$0.34& 11.07$\pm$0.33& 11.77$\pm$0.35& 10.47$\pm$0.32& 11.79$\pm$0.35 \\
6300.30 [O {\sc i}]             &  0.61$\pm$0.04&  0.89$\pm$0.04&  1.21$\pm$0.04&  0.60$\pm$0.04&  0.60$\pm$0.03 \\
6312.10 [S {\sc iii}]           &  0.63$\pm$0.04&  0.78$\pm$0.04&  0.50$\pm$0.03&  0.94$\pm$0.05&  0.87$\pm$0.04 \\
6363.80 [O {\sc i}]             &  0.18$\pm$0.03&  0.30$\pm$0.03&  0.36$\pm$0.03&  0.29$\pm$0.05&  0.13$\pm$0.02 \\
6548.10 [N {\sc ii}]            &...~~~~~       &...~~~~~       &  1.65$\pm$0.06&  0.47$\pm$0.04&...~~~~~       \\
6562.80 H$\alpha$               &276.51$\pm$8.57&277.59$\pm$8.56&278.70$\pm$8.62&283.14$\pm$8.80&280.70$\pm$8.67 \\
6583.40 [N {\sc ii}]            &  1.28$\pm$0.06&  1.26$\pm$0.05&  4.61$\pm$0.15&  0.93$\pm$0.05&  1.04$\pm$0.04 \\
6678.10 He {\sc i}              &  2.73$\pm$0.10&  2.84$\pm$0.09&  2.91$\pm$0.10&  3.01$\pm$0.11&  2.49$\pm$0.08 \\
6716.40 [S {\sc ii}]            &  1.42$\pm$0.06&  2.21$\pm$0.08&  1.34$\pm$0.05&  2.28$\pm$0.08&  2.07$\pm$0.07 \\
6730.80 [S {\sc ii}]            &  1.40$\pm$0.06&  2.04$\pm$0.07&  1.31$\pm$0.05&  1.90$\pm$0.08&  1.66$\pm$0.06 \\
7065.30 He {\sc i}              &  6.75$\pm$0.22&  6.79$\pm$0.22&  7.87$\pm$0.25&  2.24$\pm$0.08&  2.66$\pm$0.09 \\
7135.80 [Ar {\sc iii}]          &  1.83$\pm$0.07&  2.54$\pm$0.09&  1.03$\pm$0.05&  2.80$\pm$0.10&  3.06$\pm$0.10 \\
7281.21 He {\sc i}              &  0.69$\pm$0.05&...~~~~~       &  0.89$\pm$0.04&  0.43$\pm$0.03&  0.58$\pm$0.03 \\
7319.90 [O {\sc ii}]            &  0.38$\pm$0.04&  0.55$\pm$0.03&  1.13$\pm$0.04&  0.45$\pm$0.03&  0.36$\pm$0.03 \\
7330.20 [O {\sc ii}]            &  0.36$\pm$0.04&  0.55$\pm$0.03&  0.87$\pm$0.04&  0.38$\pm$0.04&  0.33$\pm$0.03 \\
7751.12 [Ar {\sc iii}]          &  0.55$\pm$0.05&  0.68$\pm$0.03&  0.33$\pm$0.02&  0.45$\pm$0.03&  0.73$\pm$0.03 \\
9069.00 [S {\sc iii}]           &  2.89$\pm$0.15&  4.03$\pm$0.16&  2.05$\pm$0.10&...~~~~~       &  2.51$\pm$0.12 \\
9530.60 [S {\sc iii}]           &...~~~~~       &  9.56$\pm$0.38&...~~~~~       &  6.69$\pm$0.29&...~~~~~        \\
$C$(H$\beta$)$^{\rm a}$                  &  0.30$\pm$0.04&  0.25$\pm$0.04&  0.31$\pm$0.04&  0.65$\pm$0.04&  0.29$\pm$0.04 \\
$F$(H$\beta$)$^{\rm b}$           & 26.74$\pm$0.59& 86.34$\pm$0.13& 44.38$\pm$0.08& 21.51$\pm$0.06& 47.07$\pm$0.09 \\
EW(H$\beta$)$^{\rm c}$            &347.00$\pm$1.00&438.50$\pm$0.63&519.40$\pm$0.90&494.00$\pm$1.40&406.60$\pm$0.78 \\
\hline
  \end{tabular}

\hbox{$^{\rm a}$Extinction coefficient. For $R(V)$ = 3.1 it is linked to
selective extinction by the relation $C$(H$\beta$) = 1.47$\times$$E(B-V)$.}

\hbox{$^{\rm b}$Observed flux in 10$^{-16}$ erg s$^{-1}$ cm$^{-2}$.}

\hbox{$^{\rm c}$Equivalent width in \AA.}

  \end{table*}

%% file: tab5.tex
\begin{table*}
\caption{Ionic and total element abundances \label{tab5}}
\begin{tabular}{lccccc} \hline
Property                           &J0159+0751          &J1032+4919          &J1205+4551          &J1355+4651          &J1608+3528 \\ \hline
$T_{\rm e}$(O {\sc iii}), K          &20900$\pm$500       &19200$\pm$400       &19000$\pm$400       &19400$\pm$400       &17200$\pm$300 \\
$T_{\rm e}$(O {\sc ii}), K           &15600$\pm$300       &15500$\pm$300       &15500$\pm$300       &15600$\pm$300       &15000$\pm$300 \\
$T_{\rm e}$(Ar {\sc iii}), K         &18800$\pm$400       &17700$\pm$300       &17900$\pm$300       &18100$\pm$400       &15900$\pm$300 \\
$N_{\rm e}$(S {\sc ii}), cm$^{-3}$    &640$\pm$170         &470$\pm$110         &620$\pm$140         &260$\pm$100         &190$\pm$~80 \\ \\
O$^+$/H$^+$$\times$10$^5$           &0.138$\pm$0.009     &0.239$\pm$0.014     &0.160$\pm$0.010     &0.201$\pm$0.016     &0.155$\pm$0.009 \\
O$^{2+}$/H$^+$$\times$10$^5$         &3.304$\pm$0.190     &4.154$\pm$0.226     &2.631$\pm$0.144     &3.434$\pm$0.195     &5.836$\pm$0.304 \\
O$^{3+}$/H$^+$$\times$10$^6$         &0.598$\pm$0.051     &0.545$\pm$0.043     &0.943$\pm$0.066     &0.588$\pm$0.053     &1.393$\pm$0.104 \\
O/H$\times$10$^5$                   &3.502$\pm$0.190     &4.448$\pm$0.227     &2.886$\pm$0.145     &3.697$\pm$0.196     &6.131$\pm$0.304 \\
12+log(O/H)                         &7.544$\pm$0.024     &7.648$\pm$0.022     &7.460$\pm$0.022     &7.567$\pm$0.023     &7.788$\pm$0.022 \\ \\
N$^+$/H$^+$$\times$10$^6$           &0.087$\pm$0.005     &0.087$\pm$0.004     &0.320$\pm$0.014     &0.064$\pm$0.003     &0.076$\pm$0.003  \\
$ICF$(N)                            &  22.32             &16.43               &16.06               &16.34               &33.36   \\
N/H$\times$10$^6$                   &1.952$\pm$0.115     &1.430$\pm$0.073     &5.139$\pm$0.245     &1.049$\pm$0.062     &2.546$\pm$0.132   \\
log(N/O)                            &$-$1.254$\pm$0.035~~~&$-$1.493$\pm$0.031~~~&$-$0.749$\pm$0.030~~~&$-$1.547$\pm$0.035~~~&$-$1.382$\pm$0.031~~~\\ \\
Ne$^{2+}$/H$^+$$\times$10$^5$        &0.649$\pm$0.037     &0.751$\pm$0.042     &0.345$\pm$0.020     &0.674$\pm$0.040     &1.102$\pm$0.061 \\
$ICF$(Ne)                            &   1.02             & 1.02               &1.04                &1.03                &1.00 \\
Ne/H$\times$10$^5$                  &0.664$\pm$0.040     &0.769$\pm$0.045     &0.358$\pm$0.021     &0.694$\pm$0.043     &1.107$\pm$0.064 \\
log(Ne/O)                           &$-$0.722$\pm$0.035~~~&$-$0.762$\pm$0.034~~~&$-$0.907$\pm$0.033~~~&$-$0.726$\pm$0.036~~~&$-$0.743$\pm$0.033~~~\\ \\
S$^+$/H$^+$$\times$10$^6$           &0.027$\pm$0.001     &0.040$\pm$0.002     &0.026$\pm$0.001     &0.039$\pm$0.002     &0.037$\pm$0.001 \\
S$^{2+}$/H$^+$$\times$10$^6$         &0.171$\pm$0.014     &0.250$\pm$0.016     &0.152$\pm$0.011     &0.285$\pm$0.019     &0.379$\pm$0.022 \\
$ICF$(S)                            &   2.86             & 2.06               &2.64                &2.18                &3.91  \\
S/H$\times$10$^6$                   &0.567$\pm$0.041     &0.600$\pm$0.033     &0.470$\pm$0.029     &0.704$\pm$0.042     &1.625$\pm$0.088 \\
log(S/O)                            &$-$1.791$\pm$0.040~~~&$-$1.870$\pm$0.032~~~&$-$1.788$\pm$0.034~~~&$-$1.720$\pm$0.035~~~&$-$1.577$\pm$0.032~~~\\ \\
Cl$^{2+}$/H$^+$$\times$10$^8$        &      ...           &0.560$\pm$0.072     &    ...             &   ...              &... \\
$ICF$(Cl)                           &      ...           & 1.52               &    ...             &   ...              &... \\
Cl/H$\times$10$^8$                  &      ...           &0.853$\pm$0.110     &    ...             &   ...              &... \\
log(Cl/O)                           &      ...           &$-$3.718$\pm$0.060~~~&    ...             &   ...              &... \\ \\
Ar$^{2+}$/H$^+$$\times$10$^7$        &0.513$\pm$0.022     &0.781$\pm$0.031     &0.309$\pm$0.176     &0.828$\pm$0.033     &1.116$\pm$0.044  \\
$ICF$(Ar)                           &   2.14             &  1.79              & 1.77               & 1.79               &2.78  \\
Ar/H$\times$10$^7$                  &1.098$\pm$0.191     &1.396$\pm$0.128     &0.546$\pm$0.078     &1.479$\pm$0.185     &3.101$\pm$0.353  \\
log(Ar/O)                           &$-$2.504$\pm$0.079~~~&$-$2.503$\pm$0.046~~~&$-$2.723$\pm$0.066~~~&$-$2.398$\pm$0.062~~~&$-$2.296$\pm$0.054~~~\\ \\
$[$Fe {\sc iii}$]$ $\lambda$4658: \\
Fe$^{2+}$/H$^+$$\times$10$^6$        &      ...           &0.066$\pm$0.009     &0.102$\pm$0.010     &  ...               &0.054$\pm$0.007  \\
$ICF$(Fe)                           &      ...           & 25.01              &24.64               &  ...               &51.77  \\
Fe/H$\times$10$^6$                  &      ...           &1.662$\pm$0.235     &2.516$\pm$0.247     &  ...               &2.780$\pm$0.354  \\
log(Fe/O)                           &      ...           &$-$1.428$\pm$0.065~~~&$-$1.059$\pm$0.048~~~&  ...               &$-$1.344$\pm$0.060~~~\\ 
$[$O/Fe$]$                          &      ...           &   0.158$\pm$0.065   &$-$0.211$\pm$0.048~~~&  ...               & \,0.073$\pm$0.060\\ 
\\
$[$Fe {\sc iii}$]$ $\lambda$4988: \\
Fe$^{2+}$/H$^+$$\times$10$^6$        &      ...           &0.100$\pm$0.009     &  ...               &  ...               &...  \\
$ICF$(Fe)                           &      ...           & 25.01              &  ...               &  ...               &...  \\
Fe/H$\times$10$^6$                  &      ...           &2.495$\pm$0.236     &  ...               &  ...               &...  \\
log(Fe/O)                           &      ...           &$-$1.251$\pm$0.047~~~&  ...               &  ...               &...  \\
$[$O/Fe$]$                          &      ...           &$-$0.169$\pm$0.047~~~&  ...               &  ...               &...  \\
\hline
  \end{tabular}
  \end{table*}

%% file: tab6.tex
\begin{table*}
\caption{Broad H$\alpha$ emission \label{tab6}}
\begin{tabular}{lcccc} \hline
Name      &$F$(broad)/$F$(total)&FWHM \\
          &    (\%)               &(km s$^{-1}$) \\ \hline
J0159+0751&    1.9              &  1850 \\
J1032+4919&    1.4              &  1670 \\
J1205+4551&    2.6              &  2040 \\
J1355+4651&    0.5              &  1750 \\
J1608+3528&    0.7              &  1680 \\
\hline
  \end{tabular}
  \end{table*}

%% file: tab7.tex
\begin{table*}
\caption{He {\sc i} emission-line ratios, electron density
$N_{\rm e}$(He {\sc i}) and optical depth $\tau$(3889) \label{tab7}}
\begin{tabular}{lccccccccc} \hline
Name      &$N_{\rm e}$(He {\sc i})&$\tau$(He {\sc i} $\lambda$3889)&\multicolumn{3}{c}{$I$(He {\sc i} $\lambda$3889)/$I$(He {\sc i} $\lambda$6678)}&&
\multicolumn{3}{c}{$I$(He {\sc i} $\lambda$7065)/$I$(He {\sc i} $\lambda$6678)} \\ \cline{4-6} \cline{8-10}
          &(cm$^{-3}$)&  &obs.&cal.&low-density&&obs.&cal.&low-density \\
          &          &  &    &    &case B&&    &    &case B \\ \hline
J0159+0751&2246\,\,\,& 0.8      &2.86&2.79&3.91&&2.47&2.47&0.92 \\
J1032+4919& 695      & 7.2      &2.04&2.04&3.78&&2.39&2.39&0.87 \\
J1205+4551& 595      &10.5\,\,\,&1.48&1.48&3.76&&2.70&2.70&0.87 \\
J1355+4651& 314      & 0.0      &3.39&3.51&3.80&&0.74&1.01&0.88 \\
J1608+3528& 399      & 0.0      &3.49&3.42&3.61&&1.07&0.98&0.82 \\
\hline
  \end{tabular}
  \end{table*}